\documentclass[aps, twocolumn,superscriptaddress, 
amsmath,amssymb,reprint,numbers,noeprint, floatfix,longbibliography]{revtex4-1}

\usepackage{graphicx}
\graphicspath{
      {figures/}
      {../figures/}
    }
\usepackage{dcolumn}
\usepackage{bm}

\usepackage{float}

\usepackage{physics}

\usepackage{color}
\definecolor{red}{rgb}{1,0,0}
\definecolor{blue}{rgb}{0,0,1}

\newcommand{\SIdevFab}{1}

\newcommand{\SIcmt}{4}
\newcommand{\SIfitting}{5}
\newcommand{\SIoptSim}{6}
\newcommand{\SIoptHighQ}{7}
\newcommand{\SIbvd}{8}
\newcommand{\SItimeGate}{9}
\newcommand{\SIheterodyne}{10}
\newcommand{\SIdft}{11}

\begin{document}

\title{Silicon-lattice-matched boron-doped gallium phosphide: A scalable acousto-optic platform}

\author{Nicholas S. Yama}
\thanks{nsyama@uw.edu, These two authors contributed equally}
\affiliation{University of Washington, Electrical and Computer Engineering Department, Seattle, WA, 98105, USA}%

\author{I-Tung Chen}
\thanks{These two authors contributed equally}
\affiliation{University of Washington, Electrical and Computer Engineering Department, Seattle, WA, 98105, USA}%

\author{Srivatsa Chakravarthi}
\affiliation{University of Washington, Physics Department, Seattle, WA, 98105, USA}%

\author{Bingzhao Li}
\affiliation{University of Washington, Electrical and Computer Engineering Department, Seattle, WA, 98105, USA}%

\author{Christian Pederson}
\affiliation{University of Washington, Physics Department, Seattle, WA, 98105, USA}%

\author{Bethany E. Matthews}
\affiliation{Energy and Environment Directorate, Pacific Northwest National Laboratory, Richland, Washington 99352, USA}%

\author{Steven R. Spurgeon}
\affiliation{Energy and Environment Directorate, Pacific Northwest National Laboratory, Richland, Washington 99352, USA}%
\affiliation{University of Washington, Physics Department, Seattle, WA, 98105, USA}%

\author{Daniel E. Perea}
\affiliation{Earth and Biological Sciences Directorate, Environmental Molecular Sciences Laboratory, Pacific Northwest National Laboratory, Richland, Washington 99352, USA}%

\author{Mark G. Wirth}
\affiliation{Earth and Biological Sciences Directorate, Environmental Molecular Sciences Laboratory, Pacific Northwest National Laboratory, Richland, Washington 99352, USA}%

\author{Peter V. Sushko}
\affiliation{Physical and Computational Sciences Directorate, Pacific Northwest National Laboratory, Richland, Washington 99352, USA}%

\author{Mo Li}
\affiliation{University of Washington, Electrical and Computer Engineering Department, Seattle, WA, 98105, USA}
\affiliation{University of Washington, Physics Department, Seattle, WA, 98105, USA}%

\author{Kai-Mei C. Fu}
\affiliation{University of Washington, Electrical and Computer Engineering Department, Seattle, WA, 98105, USA}
\affiliation{University of Washington, Physics Department, Seattle, WA, 98105, USA}
\affiliation{Physical Sciences Division, Pacific Northwest National Laboratory, Richland, Washington 99352, USA}%

\date{\today}

\begin{abstract}
    The compact size, scalability, and strongly confined fields in integrated photonic devices enable new functionalities in photonic networking and information processing, both classical and quantum. 
    Gallium phosphide (GaP) is a promising material for active integrated photonics due to its high refractive index, wide band gap, strong nonlinear properties, and large acousto-optic figure of merit. 
    In this work we demonstrate that silicon-lattice-matched boron-doped GaP (BGaP), grown at the 12-inch wafer scale, provides similar functionalities as GaP. 
    BGaP optical resonators exhibit intrinsic quality factors exceeding 25{,}000 and 200{,}000 at visible and telecom wavelengths respectively. 
    We further demonstrate the electromechanical generation of low-loss acoustic waves and an integrated acousto-optic (AO) modulator.
    High-resolution spatial and compositional mapping, combined with \textit{ab initio} calculations indicate two candidates for the excess optical loss in the visible band: the silicon-GaP interface and boron dimers.
    These results demonstrate the promise of the BGaP material platform for the development of scalable AO technologies at telecom and provide potential pathways toward higher performance at shorter wavelengths.
\end{abstract}

\maketitle

On-chip photonic devices are central to quantum information technologies in individual platforms such as trapped ions \cite{brown2021NatRevMat_ions} and quantum emitters in solid-state \cite{bradac2020NatComm_diaDefects, lukin2020prxq_sic}, as well as in interconnects between systems for quantum networks \cite{awschalom2021prxq_quics}. 
The integration of such photonic devices with mechanical degrees of freedom in an acousto-optic (AO) platform enables not only the control of qubits \cite{maity2020natComm_sivAcousticCtrl, shandilya2021natPhys_optomech} and microwave-to-optical transduction \cite{jiang2020natComm_transduction,honl2022natComm_gapTransduction}, but also access to the rich physics of classical and quantum cavity optomechanics \cite{Aspelmeyer2014RevModPhys_cavOptoMech}.
The development of an AO platform requires a versatile material with strong nonlinear properties and a scalable growth and fabrication process.

Gallium phosphide (GaP) has shown significant promise in integrated photonics due to its large band gap (2.24\,eV), high refractive index ($n_{\text{GaP}} = 3.34$ at 600\,nm \cite{bond1965jApplPhys_gapindex}), and large second-order nonlinearity ($\chi^{(2)}\sim 110$\,pm/V \cite{dal1996prb_gapChi2}).
This rare combination of a wide transparency window and large refractive index makes GaP uniquely attractive for integration with color centers embedded in high-index hosts such as diamond ($n_{\text{dia}} = 2.4$) and silicon carbide ($n_{\text{SiC}} = 2.6$); both of which, are incompatible with leading nonlinear materials such as lithium niobate (LiNbO$_3$, $n_{\text{LiNbO$_3$}} = 2.3$).
There are numerous demonstrations of the hybrid GaP-on-diamond photonics platform including high-efficiency photon extractors \cite{chakravarthi2020optica_inverseDesign}, micro-disk cavity-coupled defects \cite{gould2015josab_gap, gould2016prappl_gap, schmidgall2018nanolett_stark}, and small-mode-volume photon-defect interfaces \cite{chakravarthi2023nanolett_stamp}.
Likewise, the strong nonlinear properties have been leveraged to perform high-efficiency visible-to-telecom conversion \cite{logan2023optexp_triply} and frequency comb generation \cite{wilson2020natphoton_nonlinearGaP}.

In addition to its excellent optical properties, GaP also has strong acoustic properties which makes it particularly attractive for AO devices.
In particular, the slow speed of sound ($v_{s,\text{GaP}} \approx 4100$\,m/s \cite{Weil1968JAP_gapVelocity}) and large photoelastic coefficient ($p_{11}=-0.24$ \cite{Mytsyk2015_gapphotoelastic}; compare to LiNbO$_3$ $p_{13} = 0.172$ \cite{Weis1985_LNOphotelastic}) enables efficient guiding of acoustic waves and a large AO interaction strength.
These properties combine to give GaP one of the largest AO figures-of-merit of $\mathcal M_{\text{GaP}} = 44.6\times10^{-15}\,\mathrm{s^3 kg^{-1}}$ which is about 1.5 times that of LiNbO$_3$ ($\mathcal M_{\text{LiNbO$_3$}} = 29.2\times10^{-15}\,\mathrm{s^3 kg^{-1}}$) \cite{Dixon1967_FOM}.
These favorable properties have been leveraged in cavity-optomechanical systems \cite{schneider2019optica_gapOptomech, mitchell2014applPhysLett_gapOptomech, stockhill2019prl_gapOptomech} and microwave-to-optical frequency transduction \cite{honl2022natComm_gapTransduction}.
Additionally, the large contrast with the speed of sound in diamond ($v_{s,\text{dia}} \sim 12000$\,m/s \cite{WANG2004_diamond_aco}) enables efficient guiding of acoustic waves which can be leveraged for hybrid cavity optomechanics \cite{ma2023arxiv_gapDiaOptoMech}.

Here we demonstrate that commercially available boron-doped GaP (BGaP), with doping-enabled lattice-matched growth on silicon \cite{nasp2008jCrystalGrowth_latticematch,nasp2011jApplPhys_BGaP1,nasp2011jApplPhys_BGaP2,nasp2017jApplPhys_BGaP3}, is a scalable alternative to GaP for integrated AO devices and can be readily adapted to a variety of qubit systems and architectures.
We demonstrate techniques for both hybrid/suspended device fabrication and characterize the material's optical loss through visible- and telecom-wavelength photonic resonators.
Our analysis shows material-limited optical quality factors at visible wavelengths (650--800\,nm) of $Q_{i,\text{vis}} \approx 25{,}000$ and establishes a lower-bound for the material limit at telecom wavelengths (1530--1565\,nm) to be $Q_{i,\text{tel}} \gtrsim 200{,}000$.
In the acoustic domain, we demonstrate the electro-mechanical generation of acoustic waves, characterize acoustic loss, and implement an on-chip AO frequency shifter (AOFS).
High-resolution imaging and compositional mapping, combined with theoretical calculations, point to two potential sources of loss at visible wavelengths: the Si-GaP interface and boron dimers; thus providing a pathway to further enhance performance.

\section{Results and discussion}
\subsection{BGaP-on-silicon}
\begin{figure}[t]
    \centering
    \includegraphics{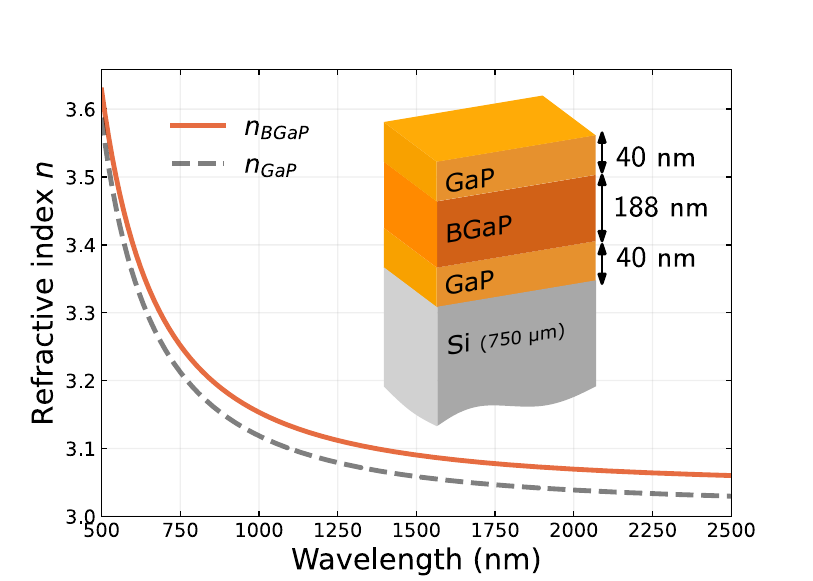}
    \caption{\textbf{BGaP-on-silicon.}
    Schematic of the GaP-BGaP-GaP structure as grown on the silicon substrate and real part of the refractive index $n$ of BGaP and GaP as determined by variable angle spectroscopic ellipsometry.
    }
    \label{fig:bgap_on_si}
\end{figure}

Commercial BGaP (NAsP$_{\text{III/V}}$ GmbH) is grown on a silicon (Si) substrate in a symmetric GaP-BGaP-GaP heterostructure (40\,nm-188\,nm-40\,nm) as shown schematically in Fig.~\ref{fig:bgap_on_si}a.
The incorporation of boron in the BGaP layer at around 2.77\% is used to facilitate lattice matching to the Si growth substrate~\cite{nasp2008jCrystalGrowth_latticematch}.

Variable-angle spectroscopic ellipsometry in the GaP transparency window (500\,nm--2500\,nm) was performed to determine the refractive index (Fig.~\ref{fig:bgap_on_si}).
The refractive index of the BGaP layer is found to be slightly larger than intrinsic GaP with $\Delta n \approx 0.04$, which is consistent with previous measurements at single wavelengths \cite{nasp2017jApplPhys_BGaP3}.
Loss within in the band gap is below the ellipsometry detection limit.

\subsection{Visible-wavelength device characterization}
\begin{figure*}[t]
    \centering
    \includegraphics[width=\textwidth]{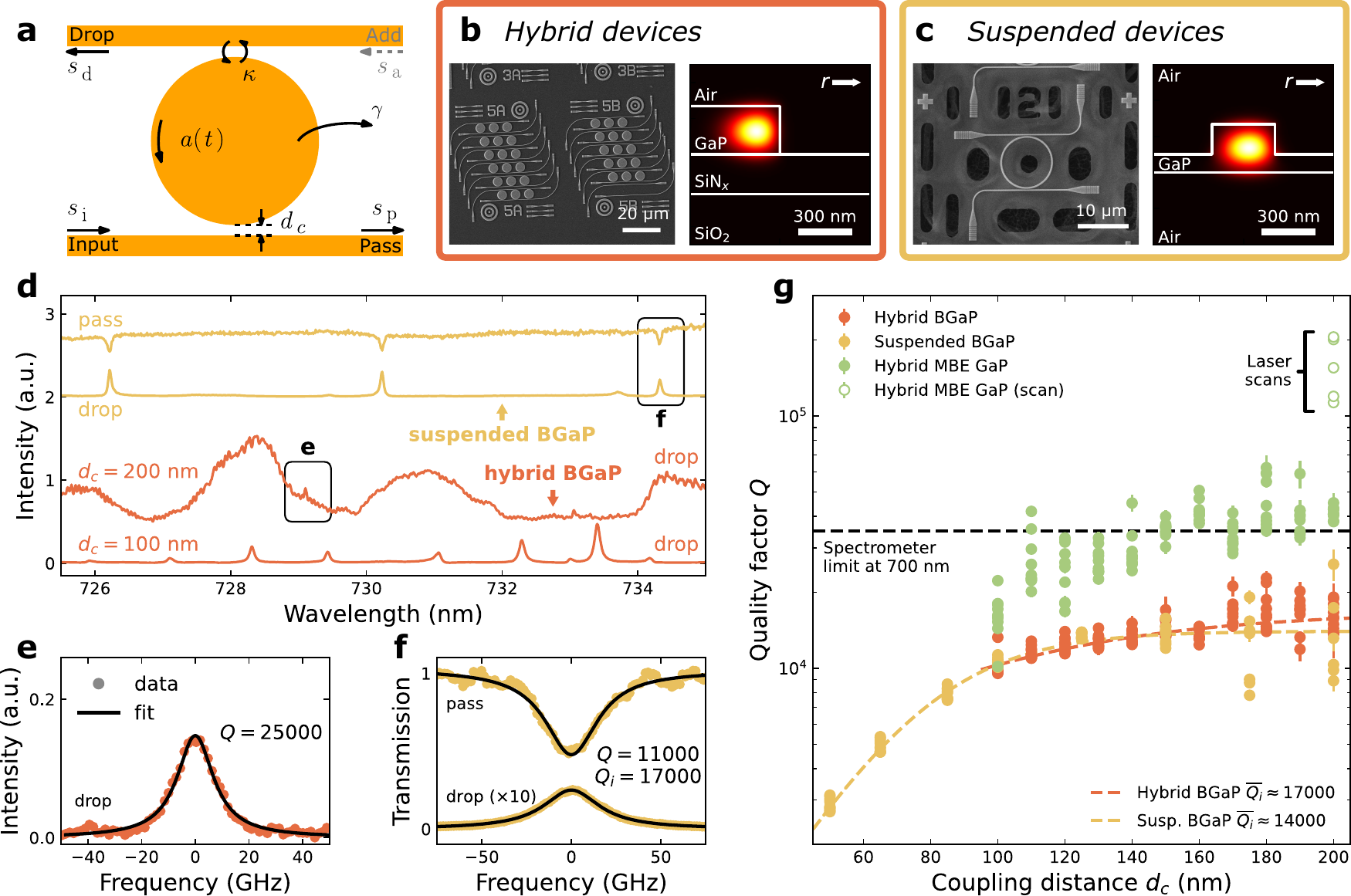}
    \caption{\textbf{Visible-wavelength device characterization.} 
    \textbf{(a)} Schematic of the add-drop resonator configuration.
    \textbf{(b,c)} SEM images and simulated mode profiles for the \textbf{(b)} hybrid and \textbf{(c)} suspended device geometries respectively. 
    \textbf{(d)} Representative transmission spectra of the devices at different ports and coupling distances $d_c$.
    Red curves correspond to the hybrid devices throughout whereas the yellow curves correspond to the suspended devices.
    The labeled resonances correspond to the high-resolution laser scan measurements in \textbf{(e)} and \textbf{(f)}.
    \textbf{(g)} Resonance quality factors as a function of coupling distance $d_c$ for both geometries.
    The green circles represent MBE-grown un-doped GaP devices with the same hybrid geometry design showing quality factors exceeding $10^5$ in scanning laser measurements.
    The spectrometer resolution limit corresponds to a $Q\approx35{,}000$.
    The hybrid and symmetric data is fitted assuming an exponential dependence on $d_c$ of $Q_c$ (SI.\SIfitting) yielding average intrinsic quality factors of $\overline{Q_i}\approx 16{,}000$.
    }
    \label{fig:visible_devices}
\end{figure*}

To characterize the photonic loss of the BGaP material in the visible wavelength range we fabricated nanophotonic disk and ring resonators.
The resonators are coupled evanescently to symmetrically positioned linear waveguides in an add-drop configuration (Fig. \ref{fig:visible_devices}a).

This geometry is modeled through temporal coupled-mode theory (CMT) \cite{manolatou1999ieee_addDropCMT, li2016lasRev_backscattering} (SI.\SIcmt).
For a resonance with angular frequency $\omega_0$, intrinsic loss rate $\gamma$, and coupling rate $\kappa$ to each waveguide (total coupling is $2\kappa$), the total (loaded) quality factor $Q$ given by
\begin{equation}
    \frac{1}{Q} = \frac{1}{Q_i} + \frac{2}{Q_{c}}
\end{equation}
where $Q_i \equiv \omega_0/\gamma$ and $Q_{c}\equiv\omega_0/\kappa$ are the intrinsic and coupling quality factors respectively.
The intrinsic quality factor is dependent on the radiation loss of the resonator structure, fabrication imperfections, and material absorption. 
By varying the coupling distance $d_c$ (thus modulating $Q_c$), we determine the intrinsic quality factor $Q_i$ which provides an upper bound on the material-induced loss rates.

We fabricated 5- and 10-\textmu m diameter disk and ring resonators in both hybrid BGaP-on-SiN$_x$ (Fig.~\ref{fig:visible_devices}b) and suspended (Fig.~\ref{fig:visible_devices}c) architectures respectively (details provided in the Methods, SI.\SIdevFab). 
The coupling distance $d_c$ between the resonator and waveguide is varied between 50--200\,nm to probe from the over- ($\kappa > \gamma$) to under-coupled ($\kappa \ll \gamma$) regimes.
SiN$_x$ ($n_{\mathrm{SiN_x}} = 2.0$) was used to demonstrate the potential for high-index substrate integration.
Radiation losses into the substrate are determined to be negligible (SI.\SIoptSim).

We perform broadband transmission measurements by exciting at the input port with a supercontinuum laser (600--800\,nm) and collecting at either the drop or pass ports as denoted in Fig.~\ref{fig:visible_devices}a.
The transmission signal is then collected on a spectrometer with 0.02-nm-per-pixel resolution.
Representative broadband pass and drop spectra are shown in Fig.~\ref{fig:visible_devices}d over a few free spectral ranges.

Fitted quality factors are plotted in Fig.~\ref{fig:visible_devices}g for both the hybrid BGaP-on-SiN$_x$ (red) and suspended BGaP (yellow) device geometries.
The asymptoting of the total quality factors for the BGaP devices for coupling distances exceeding $d_c \approx 120$\,nm signifies the under-coupled regime in which $Q \approx Q_i$.
Despite the significant differences in both geometry and fabrication, both the hybrid- and suspended-BGaP devices asymptote to similar values of $Q\approx 16{,}000$, suggesting an intrinsic material limitation.

This hypothesis of a material limitation in BGaP is further supported by measurements of un-doped GaP-on-SiN$_x$ devices (Fig.~\ref{fig:visible_devices}g, green) which were fabricated using similar methods.
These devices exhibited resonance $Q$'s beyond the spectrometer resolution limit.
Using scanning-laser transmission measurements (725--740\,nm), we measured quality factors as high as $200{,}000$ in the un-doped GaP devices (SI.\SIoptHighQ).
The highest measured total $Q$ for the BGaP devices was $25{,}000$ as shown in Fig.~\ref{fig:visible_devices}e,f.

\subsection{Telecom-wavelength device characterization}

\begin{figure*}[t]
    \centering
    \includegraphics[width=\textwidth]{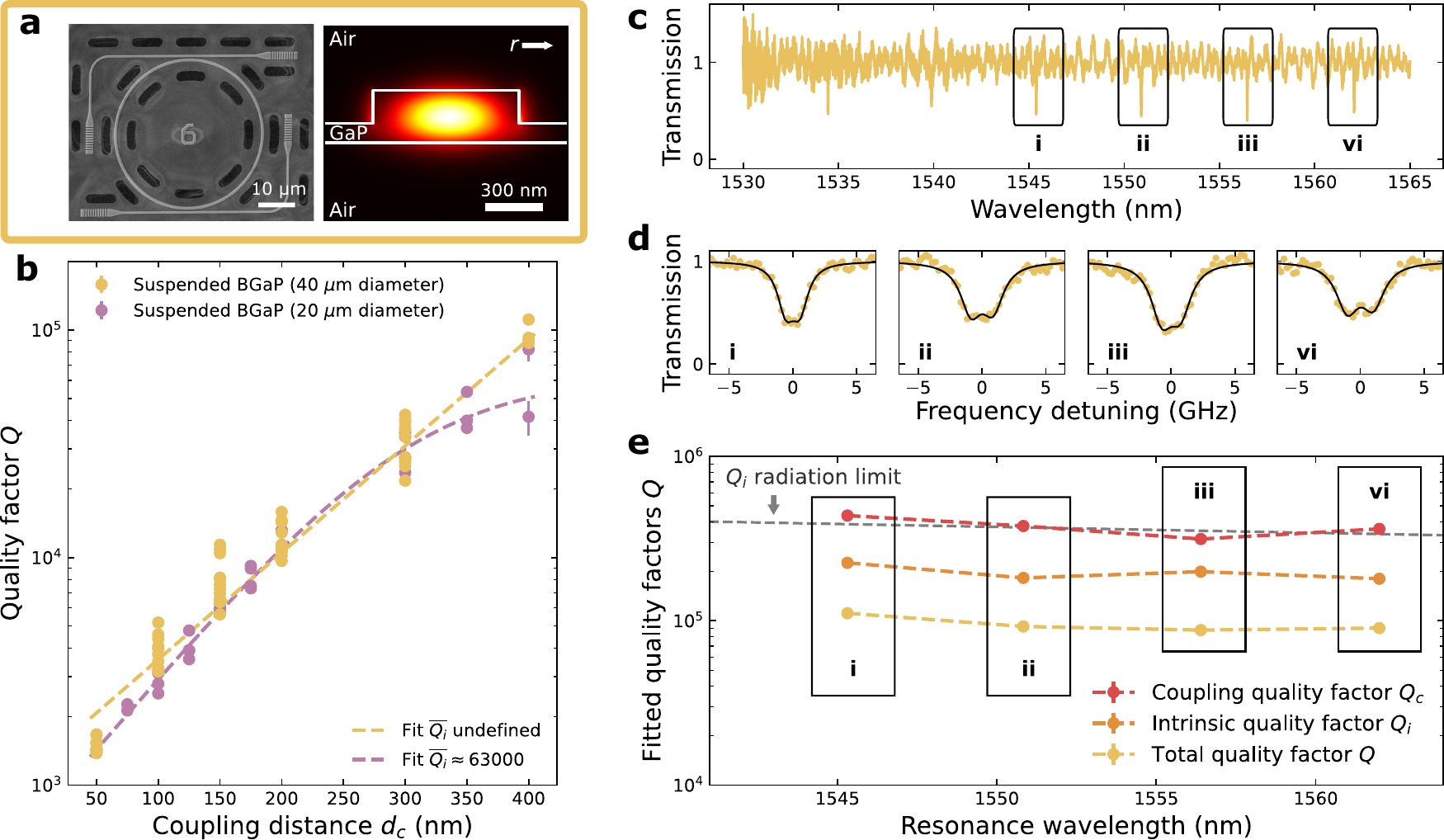}
    \caption{\textbf{Telecom-wavelength device characterization.} 
    \textbf{(a)} SEM and simulated mode profile for the 40-\textmu m diameter telecom resonator.
    \textbf{(b)} Measured total quality factors for the 40-\textmu m (yellow) and  20-\textmu m (purple) ring resonators as a function of coupling distance $d_c$.
    Asymptoting could not be observed for the 40-\textmu m ring devices and so the saturation $\overline{Q_i}$ could not be determined.
    \textbf{(c)} Pass-port transmission spectra for the most under-coupled 40-\textmu m device ($d_c = 400$\,nm).
    \textbf{(d)} High-resolution stepped sweeps of the indicated resonances in \textbf{(c)}.
    The modes are split due to back-scattering/coupling which is modeled in SI.\SIcmt.
    \textbf{(e)} Total, intrinsic, and coupling quality factors of the corresponding resonances as extracted from the fits.
    The radiation limited quality factor $Q_r$ was simulated to be around $400{,}000$.
    The intrinsic quality factor is measured to be around $Q_i \approx 200{,}000$.
    }
    \label{fig:telecom_devices}
\end{figure*}

We utilize scanning-laser transmission measurements (1530--1565\,nm) to characterize the BGaP optical quality at telecom wavelengths.
Two sets of add-drop ring resonator devices with diameters of 20 and 40\,\textmu m were fabricated in the suspended device geometry (Fig.~\ref{fig:telecom_devices}a).
The coupling distances are varied between 50--400\,nm.

Quality factors as a function of coupling distance are shown in Fig.~\ref{fig:telecom_devices}b.
Representative normalized transmission spectra for the most under-coupled 40-\textmu m device ($d_c = 400$\,nm) is shown in Fig.~\ref{fig:telecom_devices}c.
These resonances exhibit high quality factors exceeding $100{,}000$.
As a result, the fabricated coupling distances did not reach the very under-coupled regime.

We obtain estimates for the intrinsic quality factor by directly fitting the transmission spectra to the CMT model.
For these resonances, an asymmetric mode splitting is observed, which is understood to be a result of both back-scattering and back-coupling of the counter-propagating modes within the resonator \cite{li2016lasRev_backscattering}.
We derive the transmission spectra to be given by
\begin{equation}
    T_p = 
    \left|1
    -\frac{1}{2Q_c} \left(
        \frac{1 + 2|f|\cos\beta}{i\frac{\Delta + |g|}{\omega_0} + \frac{1}{Q}} + 
        \frac{1 - 2|f|\cos\beta}{i\frac{\Delta + |g|}{\omega_0} + \frac{1}{Q}}
    \right)
    \right|^2
\end{equation}
where $\Delta = \omega_0 - \omega$ is the detuning from the resonance frequency $\omega_0$, $|g|$ is the back-scattering rate, and $|f|\cos\beta$ describes the relative amplitude and phase of the back-coupling rate (see SI.\SIcmt).
If we assume the $|f|\ll1$, a fit to this model determines both the coupling and intrinsic factors. 

We determine total quality factors of $Q\approx 100{,}000$ and intrinsic quality factors $Q_i \approx 200{,}000$ (Fig.~\ref{fig:telecom_devices}e).
This analysis also reveals that the coupling quality factor $Q_c$ varies between $300{,}000$--$400{,}000$ demonstrating that these devices are near the critically coupled regime of $Q_i = Q_c/2$.
This is consistent with the lack of asymptoting observed in Fig.~\ref{fig:telecom_devices}b.

Finally, the simulated radiation-limited $Q$ for this design (Fig.~\ref{fig:telecom_devices}e, see SI.\SIoptSim) is also determined to vary between $400{,}000$--$300{,}000$ over the measurement range as a result of weaker confinement at larger wavelengths.
Since the contributions of radiation, fabrication, and material losses add in parallel to make up the intrinsic quality factor, this puts a lower bound on the material-limited intrinsic quality factor to be at least $300{,}000$--$400{,}000$ in the telecom C band.
Similar quality factors have been observed in un-doped GaP photonic devices \cite{wilson2020natphoton_nonlinearGaP} which is promising for applications \cite{logan2023optexp_triply, honl2022natComm_gapTransduction}.

\subsection{Acoustic characterization}

\begin{figure*}[t]
    \centering
    \includegraphics[width=\textwidth]{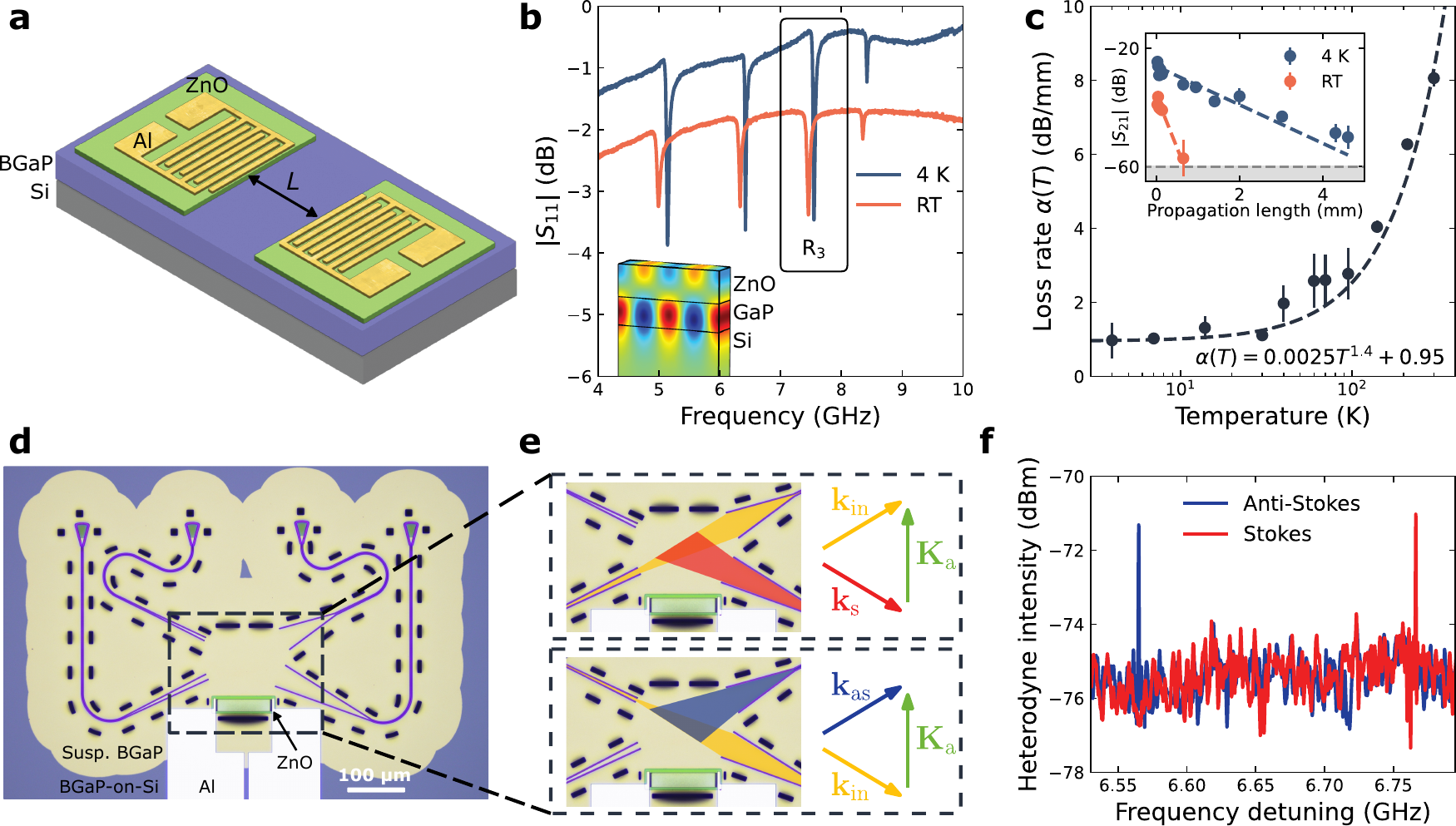}
    \caption{\textbf{BGaP acoustic devices and characterization.} 
    \textbf{(a)} A schematic of acoustic delay line device. 
    \textbf{(b)} $|S_{11}|$ of the IDT at RT and 4\,K.
    The highlighted Rayleigh mode ($R_3$, profile inset) is used for propagation loss measurements.
    \textbf{(c)} Temperature-dependent acoustic propagation loss. The inset shows the $|S_{21}|$ of the acoustic delay line at RT and 4\,K.
    \textbf{(d)} An optical image of the AOFS device.
    \textbf{(e)} Optical and acoustic propagation schematic and phase-matching conditions. 
    Here $k_{\text{in}}$ is the input optical, $K_{a}$ is the input acoustic, and $k_S$ ($k_{AS}$) is the Stokes (anti-Stokes) optical wavevectors respectively. 
    \textbf{(f)} Heterodyne measurement spectrum of the AOFS.
    }
    \label{fig:acoustics}
\end{figure*}

\begin{figure*}[t]
    \centering
    \includegraphics[width=\textwidth]{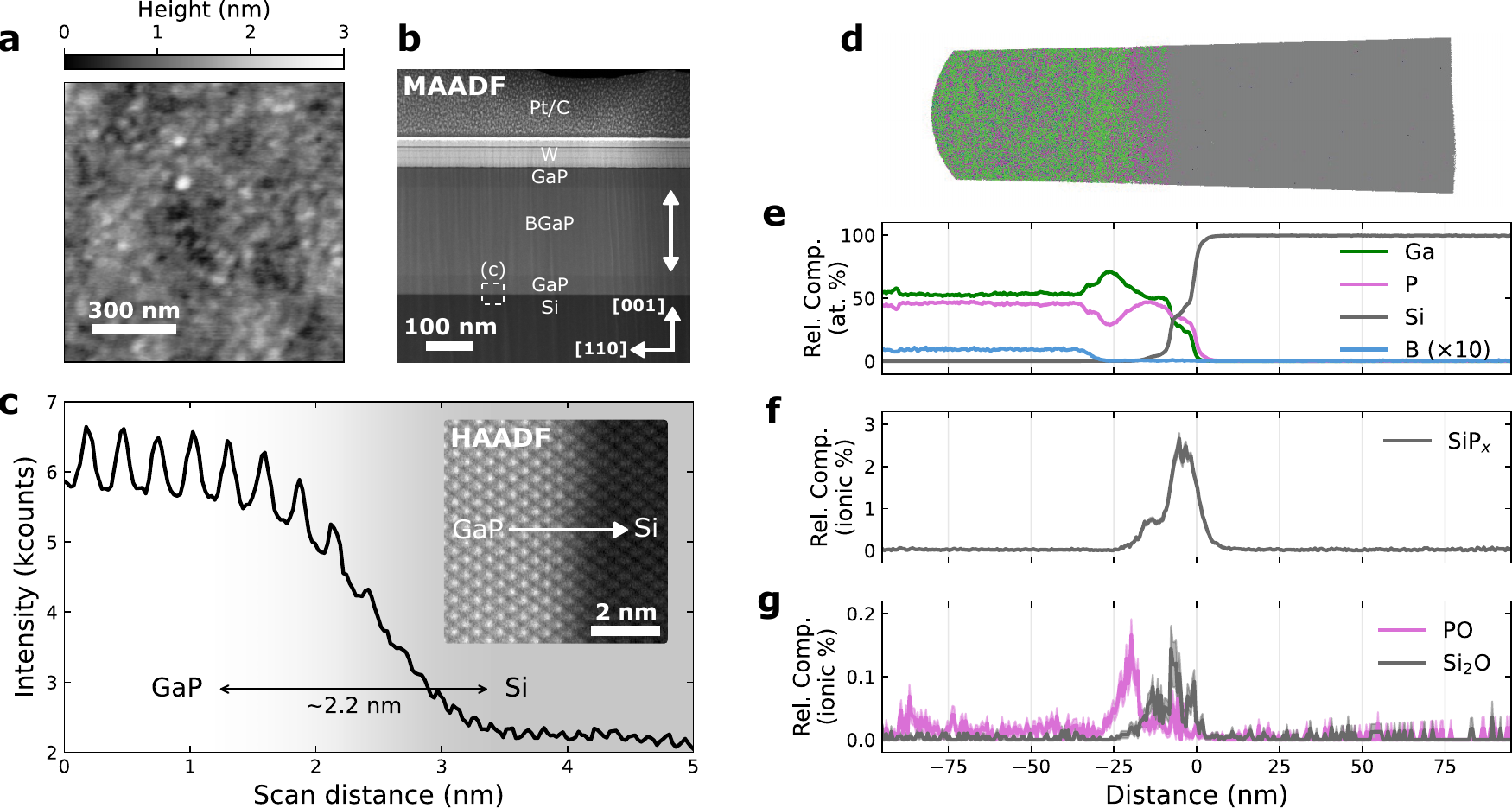}
    \caption{\textbf{High-resolution spatial mapping.} 
    \textbf{(a)} AFM image of the top GaP surface.
    \textbf{(b)} Medium-angle annular dark-field (MAADF) cross-sectional STEM image of the BGaP stack.
    Dashed box indicates the approximate region of the high-resolution image in \textbf{c}. 
    The vertical striations in these images are the result of curtaining during focused ion beam sample preparation.
    \textbf{(c)} Extracted line profile and corresponding integrated signal from the inset cross-sectional drift-corrected STEM high-angle annular dark-field (HAADF) image.
    \textbf{(d)} Reconstruction of the BGaP stack and Si substrate obtained via atom probe.
    \textbf{(e)} Relative composition of Ga, P, Si, and B species as a function of depth into the stack.
    Note that the B curve is multiplied by 10.
    \textbf{(f,g)} Relative ionic composition of \textbf{(f)} SiP$_x$ and \textbf{(g)} Si$_2$O and PO compounds respectively.
    }
    \label{fig:material_characterization}
\end{figure*}

We first demonstrate surface acoustic wave (SAW) generation in BGaP-on-silicon. 
The weak piezoelectric coefficient of GaP \cite{Nelson_GaP_piezoelectric_coef} necessitates a hybrid zinc oxide (ZnO)-BGaP-Si structure to efficiently generate SAW waves (Fig.~\ref{fig:acoustics}a). 
The high acoustic velocity contrast between BGaP and Si ($\sim 8000$\,m/s) ensures low loss into the substrate, enabling the determination of the intrinsic loss without suspending the BGaP film.

SAW modes are generated by an interdigital transducer (IDT). 
The IDT consists of 65 pairs of aluminium fingers spaced with a period of $680$\,nm, an aperture of $90$\,\textmu m, and a finger width of 165\,nm. 
Fig.~\ref{fig:acoustics}b shows a typical radio-frequency (RF) reflection spectrum $|S_{11}|$ of the IDT at room-temperature (RT) and 4\,K.
Four distinct resonances can be seen in the $S_{11}$ spectrum.
We focus on the quasi-Rayleigh mode $R_3$, as finite-element simulation (Fig.~\ref{fig:acoustics}b, inset) shows the strongest localization of acoustic energy in the BGaP layer.
We extract the electro-mechanical conversion efficiency using the Butterworth-Van Dyke model \cite{Larson2000_BVD, Li2019_BVD} yielding $5\%$ at RT. At 4K, the efficiency increases to $>80\%$ (SI.\SIbvd), which is attributed to the reduction in electrode resistivity and better impedance matching.

We investigate the acoustic wave propagation loss using acoustic delay lines (Fig.~\ref{fig:acoustics}a). 
Generated SAWs propagate through the BGaP and are transduced back into an electrical signal by the receiving IDT.
The intensity of the transmission signal is a function of the propagation distance $L$, which can be expressed as $I/I_0 = e^{-\alpha L}$, where $\alpha$ is the acoustic loss rate.
Transmission measurements are performed on twelve delay line devices with identical IDTs and lengths ranging from 0.035--4.6 mm in order to extract $\alpha$.
We utilize time-gating post-processing to remove RF crosstalk \cite{Dahmani2020_gating, Mayor_2021_timegating} and acoustic reflections (SI.\SItimeGate).
After time-gating, we extract the $R_3$ transmission $|S_{21}|$ and fit to an exponential decay (Fig.~\ref{fig:acoustics}c, inset).
We find the acoustic loss to decrease from $\alpha_{\text{RT}} = 5.6$\,dB/mm at RT to around $\alpha_{\text{4K}} = 0.9$\,dB/mm at 4\,K.

The attenuation can be attributed to both temperature-dependent thermal phonon, and temperature-independent material defect scattering channels. 
We fit the temperature dependence to $\alpha(T) = bT^n + \alpha_0$ to determine the temperature-independent acoustic loss $\alpha_0=0.95$ dB/mm (Fig.~\ref{fig:acoustics}c).
Here, $b$ is a constant, and $n$ characterizes the temperature dependence, with $n = 1.40$ giving thermal phonon scattering as the dominant loss channel \cite{Slobodnik1970_acoustic_loss}. 
The measured $\alpha_0$ is comparable to the LiNbO$_3$-on-sapphire  ($\alpha=0.7$ dB/mm at 4\,K \cite{Mayor_2021_timegating}).

\subsection{Acousto-optic integrated devices}
We fabricated an AOFS (Fig.~\ref{fig:acoustics}d) operating at telecom wavelengths on suspended BGaP which consists of four optical grating couplers and an IDT on ZnO-on-BGaP.
The (anti-)Stokes process involves the stimulated emission (absorption) of a phonon at frequency $\Omega$ by an incident photon at frequency $\omega$, resulting in a scattered (blue-)red-shifted photon at $\omega \mp \Omega$.
These processes must satisfy the phase-matching condition which results in the deflection of the scattered light into the adjacent port as illustrated in Fig.~\ref{fig:acoustics}e.
Scattering occurs at the Bragg angle $\theta_B$ satisfying
\begin{equation*}
    \sin\theta_B =\frac{\lambda}{2\Lambda n_{\text{eff}}}.   
\end{equation*}
Here $\lambda = 1.55$\,\textmu m is the optical wavelength in vacuum, $\Lambda = 640$--680\,nm is the SAW wavelength in BGaP, and $n_{\text{eff}}\approx 2.5$ is the effective index of the optical mode, yielding $\theta_B\approx 27^\circ$ and a frequency shift by $\Omega/2\pi = 6.65$\,GHz.

To resolve the frequency-shifted signal, we use a heterodyne measurement scheme with reference signal shifted by $\delta/2\pi = 102.9$ MHz (Fig.~\ref{fig:acoustics}f, see SI.\SIheterodyne).
The anti-Stokes/Stokes signal (at $(\Omega -\delta)/2\pi=6.55$ \,GHz and $(\Omega +\delta)/2\pi=6.75$\,GHz respectively) results from the beating of the $\omega\pm\Omega$ scattered light with the reference signal at $\omega + \delta$. 
These results indicate the potential for the integration of BGaP AO devices on-chip.

\begin{figure*}[t]
    \centering
    \includegraphics{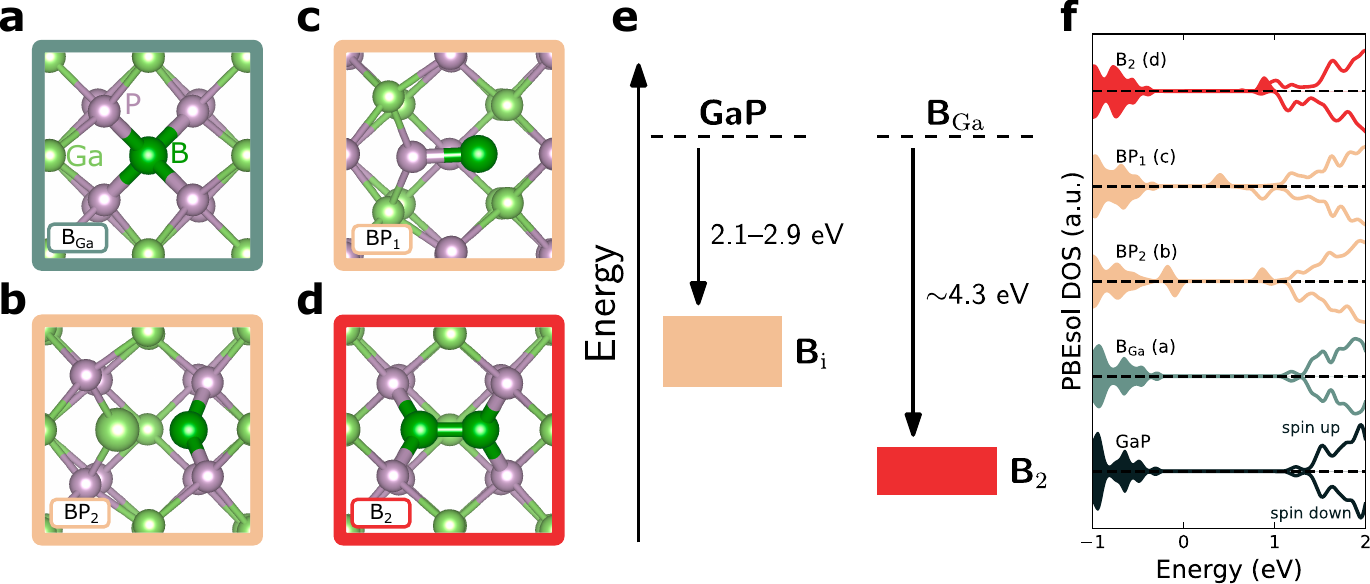}
    \caption{\textbf{Interstitial boron defects.} 
    \textbf{(a--d)} Pictorial representation of \textbf{(a)} B$_{\mathrm{Ga}}$ substitutional, \textbf{(b)} BP$_2$ interstitial, \textbf{(c)} BP$_1$ interstitial, and \textbf{(d)} B$_2$ dimer defects.
    \textbf{(e)} Energy gain due to trapping B atoms at the Ga and P lattice sites forming interstitial (B$_i$) defects (b,c) and at the substitutional B$_{\text Ga}$ defect forming B$_2$ dimers (d).
    \textbf{(f)} Spin-resolved, one-electron density of states for select configurations computed with the PBEsol functional (64-atom supercells). The calculated band energies are convolved with a Gaussian-type function of width of 0.1\,eV.
    The shaded and non-shaded regions correspond to the occupied and virtual states, respectively.    }
    \label{fig:dft}
\end{figure*}

\subsection{High-resolution spatial mapping}
In order to determine the source of the increased optical loss at visible wavelengths, we perform high-resolution spatial characterization of the BGaP.
Atomic force microscopy shows that the top surface has high surface quality with a roughness of $298 \pm 17$\,pm(Fig.~\ref{fig:material_characterization}a). 
We further examined the heterostructure quality using high-resolution scanning transmission electron microscopy (STEM).
Fig.~\ref{fig:material_characterization}b shows a cross-sectional medium-angle annular dark-field (MAADF) STEM image, revealing an overall thickness of approximately 266\,nm with smooth and defect free device layers. 
MAADF is sensitive to strain and a clear interface between the layers is present. 
Fig.~\ref{fig:material_characterization}c shows a STEM high-angle annular dark-field (HAADF) image and integrated scan, where the overall contrast is proportional to the average atomic number $Z^{\sim1.7}$.
In this mode, the film appears distinct, but uniform in intensity relative to the Si substrate, indicating its purity.
High-resolution HAADF analysis of the GaP/Si interface reveals intermixing on the order of approximately 2.2\,nm.

To characterize the composition profile across the GaP/Si interface we perform atom probe tomography (APT). Fig.~\ref{fig:material_characterization}d shows the 3-D distribution of Ga, P, and Si atoms (green, purple, and grey, respectively), with a total of over 13M counts. 
To quantify the composition across interface, a one-dimensional proximity histogram \cite{hellman2002_apt} composition profile is shown in Figs.~\ref{fig:material_characterization}e,f. 
The distance zero represents the position of a 0.85\,at.\% isoconcentration surface, which was chosen to align to the 80\% relative Si composition (Fig.~\ref{fig:material_characterization}e). 
Negative distances span the composition within the GaP material, while positive distances span within the Si substrate.
From about $-$90\,nm, the Ga, P, and B exhibit a relatively constant composition, before the B content drops to background levels around $-$35\,nm.
Between $-$35\,nm and 0\,nm, several unintentional impurity species are observed in the form of PO and Si$_2$O (Fig. 5g). 
Additionally, species consistent with SiP$_x$ are found to extend beyond the Si substrate into the GaP film over a 20\,nm distance (Fig.~\ref{fig:material_characterization}f).

Both STEM and APT point to the GaP/Si interface as a potential source of loss.
APT is used to determine the chemical composition near the interface while STEM enables precise estimates of the spatial distribution.
From APT, the observed silicon diffusion and oxygen contamination in the GaP may result in additional scattering/absorptive losses.
STEM reveals this silicon diffusion to occur at nanometer scales, which is relevant to high-$Q$ resonator modes. 
Such roughness would be exacerbated by fabrication processes like the XeF$_2$ vapor etch, which may react with the silicon to form additional contaminants or increase surface damage.
Both absorption and roughness-induced scattering are expected to more strongly affect visible-wavelength photons.

\subsection{Potential role of point defects in absorption}
In addition to the GaP/Si interface, absorption could stem from point defects, which are not detectable in STEM or APT. 
Given the large concentration of boron dopants, we examine whether boron-related defects can give rise to gap states that result in additional absorption. 
The \textit{ab initio} calculations were performed using the PBEsol density functional which, while known to underestimate band gaps, allows us to model realistic defect concentrations. 
 
We find that excess boron can form three types of interstitial boron (B$_i$) configurations by binding to the lattice P atoms: BP$_{1}$ (Fig.~\ref{fig:dft}c), BP$_{2}$ (Fig.~\ref{fig:dft}b), and BP$_{3}$.
Furthermore, we found that excess boron bound to substitutional boron (B$_{\text{Ga}}$), resulting in B$_2$ dimers (Fig.~\ref{fig:dft}d), is even more stable than B$_{i}$ bound to the lattice P only.
The formation energies, calculated relative to the ideal and B$_{\text{Ga}}$-containing GaP and gas-phase B atoms, are shown in Fig.~\ref{fig:dft}e. 

The effect of boron impurities on the electronic properties is analyzed in terms of the one-electron DOS, as shown in Fig.~\ref{fig:dft}f (see SI.\SIdft). 
Our calculations show that substitutional B$_{\text{Ga}}$ produces no gap states. 
In contrast, B$_i$ defects and B$_2$ dimers give rise to a plethora of occupied and virtual gap states, which ought to manifest as sub-band gap transitions and hence additional absorption.

\subsection{Outlook}
We have demonstrated high-performance optical resonators and on-chip acousto-optic devices, fabricated with commercially available lattice-matched BGaP grown on silicon.
Measured intrinsic optical quality factors in the telecom C band exceeding 200{,}000 and low acoustic losses comparable to leading materials are extremely promising for immediate integration in both nonlinear \cite{wilson2020natphoton_nonlinearGaP,logan2023optexp_triply} and optomechanical devices \cite{mitchell2014applPhysLett_gapOptomech, schneider2019optica_gapOptomech, honl2022natComm_gapTransduction}.
Likewise, strong visible-wavelength quality factors exceeding 25{,}000 suggest good prospects for the development of small-mode-volume, defect-integrated nanophotonic resonators \cite{chakravarthi2023nanolett_stamp} in the near term.
Our investigation revealed two potential sources of loss in the visible band: imperfections in the GaP/Si interface and optically active boron defects; both of which can be addressed by modification to the growth process, such as by the inclusion of additional buffer layers (e.g. AlGaP \cite{chakravarthi2023nanolett_stamp}) and further optimization of the boron content \cite{nasp2008jCrystalGrowth_latticematch}.
Such improvements, combined with its strong AO properties and inherent scalability, would enable BGaP to be an ideal platform for the development of next-generation integrated photonic devices.

\section*{Acknowledgments}
The authors thank Darick Baker and Cameron Toskey for assistance in ellipsometry data analysis.
The authors also thank Fariba Hatami for MBE GaP growth for high-$Q$ visible wavelength devices. Optical properties and performance, STEM analysis and interpretation (B.E.M. and S.R.S.), APT analysis (D.E.P. and M.G.W.), and computational modeling (P.V.S.) were supported by the U.S. Department of Energy (DOE) Office of Science (SC) National Quantum Information Science Research Centers, Co-design Center for Quantum Advantage (C2QA), under contract number DE-SC0012704 (Basic Energy Sciences, Pacific Northwest National Laboratory (PNNL) FWP 76274). 
Acoustic properties and performance was supported by National Science Foundation Grant No. ECCS-2006103 and the Convergence Accelerator program of the National Science Foundation (Award No. ITE-2134345).  N.S.Y. was supported by the National Science Foundation Graduate Research Fellowship Program under Grant No. DGE-2140004. 
PNNL is a multiprogram national laboratory operated for the DOE by Battelle Memorial Institute under Contract No. DE-AC05-76RL0-1830. A portion of the STEM was conducted in the Radiological Microscopy Suite (RMS), located in the Radiochemical Processing Laboratory (RPL) at PNNL. Some sample preparation was performed at the Environmental Molecular Sciences Laboratory (EMSL), a national scientific user facility sponsored by the DOE's Office of Biological and Environmental Research and located at PNNL.This research used resources of the National Energy Research Scientific Computing Center (NERSC), a DOE SC User Facility supported by the SC of the U.S. DOE under Contract No. DE-AC02-05CH11231 using NERSC award BES-ERCAP0024614. Part of this work was conducted at the Washington Nanofabrication Facility/Molecular Analysis Facility, a National Nanotechnology Coordinated Infrastructure (NNCI) site at the University of Washington with partial support from the National Science Foundation via awards NNCI-1542101 and NNCI-2025489. 

\bibliography{main}

\begin{thebibliography}{10}

\bibitem{brown2021NatRevMat_ions}
K.~R. Brown, J.~Chiaverini, J.~M. Sage, and H.~H{\"a}ffner, ``Materials
  challenges for trapped-ion quantum computers,'' {\em Nature Reviews
  Materials}, vol.~6, no.~10, pp.~892--905, 2021.

\bibitem{bradac2020NatComm_diaDefects}
C.~Bradac, W.~Gao, J.~Forneris, M.~E. Trusheim, and I.~Aharonovich, ``Quantum
  nanophotonics with group iv defects in diamond,'' {\em Nature
  Communications}, vol.~10, no.~1, p.~5625, 2019.

\bibitem{lukin2020prxq_sic}
D.~M. Lukin, M.~A. Guidry, and J.~Vu\ifmmode \check{c}\else
  \v{c}\fi{}kovi\ifmmode~\acute{c}\else \'{c}\fi{}, ``Integrated quantum
  photonics with silicon carbide: Challenges and prospects,'' {\em PRX
  Quantum}, vol.~1, p.~020102, Dec 2020.

\bibitem{awschalom2021prxq_quics}
D.~Awschalom, K.~K. Berggren, H.~Bernien, S.~Bhave, L.~D. Carr, P.~Davids,
  S.~E. Economou, D.~Englund, A.~Faraon, M.~Fejer, S.~Guha, M.~V. Gustafsson,
  E.~Hu, L.~Jiang, J.~Kim, B.~Korzh, P.~Kumar, P.~G. Kwiat,
  M.~Lon\ifmmode~\check{c}\else \v{c}\fi{}ar, M.~D. Lukin, D.~A. Miller,
  C.~Monroe, S.~W. Nam, P.~Narang, J.~S. Orcutt, M.~G. Raymer, A.~H.
  Safavi-Naeini, M.~Spiropulu, K.~Srinivasan, S.~Sun, J.~Vu\ifmmode
  \check{c}\else \v{c}\fi{}kovi\ifmmode~\acute{c}\else \'{c}\fi{}, E.~Waks,
  R.~Walsworth, A.~M. Weiner, and Z.~Zhang, ``Development of quantum
  interconnects (quics) for next-generation information technologies,'' {\em
  PRX Quantum}, vol.~2, p.~017002, Feb 2021.

\bibitem{maity2020natComm_sivAcousticCtrl}
S.~Maity, L.~Shao, S.~Bogdanovi{\'c}, S.~Meesala, Y.-I. Sohn, N.~Sinclair,
  B.~Pingault, M.~Chalupnik, C.~Chia, L.~Zheng, K.~Lai, and M.~Lon{\v c}ar,
  ``Coherent acoustic control of a single silicon vacancy spin in diamond,''
  {\em Nature Communications}, vol.~11, no.~1, p.~193, 2020.

\bibitem{shandilya2021natPhys_optomech}
P.~K. Shandilya, D.~P. Lake, M.~J. Mitchell, D.~D. Sukachev, and P.~E. Barclay,
  ``Optomechanical interface between telecom photons and spin quantum memory,''
  {\em Nature Physics}, vol.~17, no.~12, pp.~1420--1425, 2021.

\bibitem{jiang2020natComm_transduction}
W.~Jiang, C.~J. Sarabalis, Y.~D. Dahmani, R.~N. Patel, F.~M. Mayor, T.~P.
  McKenna, R.~Van~Laer, and A.~H. Safavi-Naeini, ``Efficient bidirectional
  piezo-optomechanical transduction between microwave and optical frequency,''
  {\em Nature Communications}, vol.~11, no.~1, p.~1166, 2020.

\bibitem{honl2022natComm_gapTransduction}
S.~H{\"o}nl, Y.~Popoff, D.~Caimi, A.~Beccari, T.~J. Kippenberg, and P.~Seidler,
  ``Microwave-to-optical conversion with a gallium phosphide photonic crystal
  cavity,'' {\em Nature Communications}, vol.~13, no.~1, p.~2065, 2022.

\bibitem{Aspelmeyer2014RevModPhys_cavOptoMech}
M.~Aspelmeyer, T.~J. Kippenberg, and F.~Marquardt, ``Cavity optomechanics,''
  {\em Rev. Mod. Phys.}, vol.~86, pp.~1391--1452, Dec 2014.

\bibitem{bond1965jApplPhys_gapindex}
W.~Bond, ``Measurement of the refractive indices of several crystals,'' {\em
  Journal of Applied Physics}, vol.~36, no.~5, pp.~1674--1677, 1965.

\bibitem{dal1996prb_gapChi2}
A.~D. Corso, F.~Mauri, and A.~Rubio, ``Density-functional theory of the
  nonlinear optical susceptibility: Application to cubic semiconductors,'' {\em
  Phys. Rev. B}, vol.~53, pp.~15638--15642, Jun 1996.

\bibitem{chakravarthi2020optica_inverseDesign}
S.~Chakravarthi, P.~Chao, C.~Pederson, S.~Molesky, A.~Ivanov, K.~Hestroffer,
  F.~Hatami, A.~W. Rodriguez, and K.-M.~C. Fu, ``Inverse-designed photon
  extractors for optically addressable defect qubits,'' {\em Optica}, vol.~7,
  pp.~1805--1811, Dec 2020.

\bibitem{gould2015josab_gap}
M.~Gould, S.~Chakravarthi, I.~R. Christen, N.~Thomas, S.~Dadgostar, Y.~Song,
  M.~L. Lee, F.~Hatami, and K.-M.~C. Fu, ``Large-scale gap-on-diamond
  integrated photonics platform for nv center-based quantum information,'' {\em
  J. Opt. Soc. Am. B}, vol.~33, pp.~B35--B42, Mar 2016.

\bibitem{gould2016prappl_gap}
M.~Gould, E.~R. Schmidgall, S.~Dadgostar, F.~Hatami, and K.-M.~C. Fu,
  ``Efficient extraction of zero-phonon-line photons from single
  nitrogen-vacancy centers in an integrated gap-on-diamond platform,'' {\em
  Phys. Rev. Appl.}, vol.~6, p.~011001, Jul 2016.

\bibitem{schmidgall2018nanolett_stark}
E.~R. Schmidgall, S.~Chakravarthi, M.~Gould, I.~R. Christen, K.~Hestroffer,
  F.~Hatami, and K.-M.~C. Fu, ``Frequency control of single quantum emitters in
  integrated photonic circuits,'' {\em Nano Letters}, vol.~18, pp.~1175--1179,
  02 2018.

\bibitem{chakravarthi2023nanolett_stamp}
S.~Chakravarthi, N.~S. Yama, A.~Abulnaga, D.~Huang, C.~Pederson, K.~Hestroffer,
  F.~Hatami, N.~P. de~Leon, and K.-M.~C. Fu, ``Hybrid integration of gap
  photonic crystal cavities with silicon-vacancy centers in diamond by
  stamp-transfer,'' {\em Nano Letters}, vol.~23, pp.~3708--3715, 05 2023.

\bibitem{logan2023optexp_triply}
A.~D. Logan, S.~Shree, S.~Chakravarthi, N.~Yama, C.~Pederson, K.~Hestroffer,
  F.~Hatami, and K.-M.~C. Fu, ``Triply-resonant sum frequency conversion with
  gallium phosphide ring resonators,'' {\em Opt. Express}, vol.~31,
  pp.~1516--1531, Jan 2023.

\bibitem{wilson2020natphoton_nonlinearGaP}
D.~J. Wilson, K.~Schneider, S.~H{\"o}nl, M.~Anderson, Y.~Baumgartner,
  L.~Czornomaz, T.~J. Kippenberg, and P.~Seidler, ``Integrated gallium
  phosphide nonlinear photonics,'' {\em Nature Photonics}, vol.~14, no.~1,
  pp.~57--62, 2020.

\bibitem{Weil1968JAP_gapVelocity}
R.~Weil and W.~O. Groves, ``The elastic constants of gallium phosphide,'' {\em
  Journal of Applied Physics}, vol.~39, no.~9, pp.~4049--4051, 1968.

\bibitem{Mytsyk2015_gapphotoelastic}
B.~G. Mytsyk, N.~M. Demyanyshyn, and O.~M. Sakharuk, ``Elasto-optic effect
  anisotropy in gallium phosphide crystals,'' {\em Appl. Opt.}, vol.~54,
  pp.~8546--8553, Oct 2015.

\bibitem{Weis1985_LNOphotelastic}
R.~S. Weis and T.~K. Gaylord, ``Lithium niobate: Summary of physical properties
  and crystal structure,'' {\em Applied Physics A}, vol.~37, pp.~191--203,
  1985.

\bibitem{Dixon1967_FOM}
R.~W. Dixon, ``Photoelastic properties of selected materials and their
  relevance for applications to acoustic light modulators and scanners,'' {\em
  Journal of Applied Physics}, vol.~38, no.~13, pp.~5149--5153, 1967.

\bibitem{schneider2019optica_gapOptomech}
K.~Schneider, Y.~Baumgartner, S.~H\"{o}nl, P.~Welter, H.~Hahn, D.~J. Wilson,
  L.~Czornomaz, and P.~Seidler, ``Optomechanics with one-dimensional gallium
  phosphide photonic crystal cavities,'' {\em Optica}, vol.~6, pp.~577--584,
  May 2019.

\bibitem{mitchell2014applPhysLett_gapOptomech}
M.~Mitchell, A.~C. Hryciw, and P.~E. Barclay, ``Cavity optomechanics in gallium
  phosphide microdisks,'' {\em Applied Physics Letters}, vol.~104, no.~14,
  p.~141104, 2014.

\bibitem{stockhill2019prl_gapOptomech}
R.~Stockill, M.~Forsch, G.~Beaudoin, K.~Pantzas, I.~Sagnes, R.~Braive, and
  S.~Gr\"oblacher, ``Gallium phosphide as a piezoelectric platform for quantum
  optomechanics,'' {\em Phys. Rev. Lett.}, vol.~123, p.~163602, Oct 2019.

\bibitem{WANG2004_diamond_aco}
S.-F. Wang, Y.-F. Hsu, J.-C. Pu, J.~C. Sung, and L.~Hwa, ``Determination of
  acoustic wave velocities and elastic properties for diamond and other hard
  materials,'' {\em Materials Chemistry and Physics}, vol.~85, no.~2,
  pp.~432--437, 2004.

\bibitem{ma2023arxiv_gapDiaOptoMech}
X.~Ma, P.~K. Shandilya, and P.~E. Barclay, ``Semiconductor-on-diamond cavities
  for spin optomechanics,'' 2023.

\bibitem{nasp2008jCrystalGrowth_latticematch}
B.~Kunert, S.~Zinnkann, K.~Volz, and W.~Stolz, ``Monolithic integration of
  ga(nasp)/(bga)p multi-quantum well structures on (001) silicon substrate by
  movpe,'' {\em Journal of Crystal Growth}, vol.~310, no.~23, pp.~4776--4779,
  2008.
\newblock The Fourteenth International conference on Metalorganic Vapor Phase
  Epitax.

\bibitem{nasp2011jApplPhys_BGaP1}
N.~Hossain, T.~J.~C. Hosea, S.~J. Sweeney, S.~Liebich, M.~Zimprich, K.~Volz,
  B.~Kunert, and W.~Stolz, ``Band structure properties of novel
  {B}$_x${G}a$_{1-x}${P} alloys for silicon integration,'' {\em Journal of
  Applied Physics}, vol.~110, no.~6, p.~063101, 2011.

\bibitem{nasp2011jApplPhys_BGaP2}
S.~Rogowsky, M.~Baeumler, M.~Wolfer, L.~Kirste, R.~Ostendorf, J.~Wagner,
  S.~Liebich, W.~Stolz, K.~Volz, and B.~Kunert, ``Vibrational mode and
  dielectric function spectra of {BG}a{P} probed by {R}aman scattering and
  spectroscopic ellipsometry,'' {\em Journal of Applied Physics}, vol.~109,
  no.~5, p.~053504, 2011.

\bibitem{nasp2017jApplPhys_BGaP3}
M.~Volk and W.~Stolz, ``Determination of refractive index and direct bandgap of
  lattice matched {BG}a{P} and ({BG}a)({A}s{P}) materials on exact oriented
  silicon,'' {\em Journal of Applied Physics}, vol.~122, no.~23, p.~235702,
  2017.

\bibitem{manolatou1999ieee_addDropCMT}
C.~Manolatou, M.~Khan, S.~Fan, P.~Villeneuve, H.~Haus, and J.~Joannopoulos,
  ``Coupling of modes analysis of resonant channel add-drop filters,'' {\em
  IEEE Journal of Quantum Electronics}, vol.~35, no.~9, pp.~1322--1331, 1999.

\bibitem{li2016lasRev_backscattering}
A.~Li, T.~Van~Vaerenbergh, P.~De~Heyn, P.~Bienstman, and W.~Bogaerts,
  ``Backscattering in silicon microring resonators: a quantitative analysis,''
  {\em Laser \& Photonics Reviews}, vol.~10, no.~3, pp.~420--431, 2016.

\bibitem{Nelson_GaP_piezoelectric_coef}
D.~F. Nelson and E.~H. Turner, ``Electro‐optic and piezoelectric coefficients
  and refractive index of gallium phosphide,'' {\em Journal of Applied
  Physics}, vol.~39, no.~7, pp.~3337--3343, 1968.

\bibitem{Larson2000_BVD}
J.~Larson, P.~Bradley, S.~Wartenberg, and R.~Ruby, ``Modified butterworth-van
  dyke circuit for fbar resonators and automated measurement system,'' in {\em
  2000 IEEE Ultrasonics Symposium. Proceedings. An International Symposium
  (Cat. No.00CH37121)}, vol.~1, pp.~863--868 vol.1, 2000.

\bibitem{Li2019_BVD}
H.~Li, Q.~Liu, and M.~Li, ``Electromechanical brillouin scattering in
  integrated planar photonics,'' {\em APL Photonics}, vol.~4, no.~8, p.~080802,
  2019.

\bibitem{Dahmani2020_gating}
Y.~D. Dahmani, C.~J. Sarabalis, W.~Jiang, F.~M. Mayor, and A.~H. Safavi-Naeini,
  ``Piezoelectric transduction of a wavelength-scale mechanical waveguide,''
  {\em Phys. Rev. Appl.}, vol.~13, p.~024069, Feb 2020.

\bibitem{Mayor_2021_timegating}
F.~M. Mayor, W.~Jiang, C.~J. Sarabalis, T.~P. McKenna, J.~D. Witmer, and A.~H.
  Safavi-Naeini, ``Gigahertz phononic integrated circuits on thin-film lithium
  niobate on sapphire,'' {\em Phys. Rev. Appl.}, vol.~15, p.~014039, Jan 2021.

\bibitem{Slobodnik1970_acoustic_loss}
A.~J. Slobodnik, P.~H. Carr, and A.~J. Budreau, ``Microwave frequency acoustic
  surface‐wave loss mechanisms on linbo3,'' {\em Journal of Applied Physics},
  vol.~41, no.~11, pp.~4380--4387, 1970.

\bibitem{hellman2002_apt}
O.~C. Hellman, J.~A. Vandenbroucke, J.~Rüsing, D.~Isheim, and D.~N. Seidman,
  ``{Analysis of Three-dimensional Atom-probe Data by the Proximity
  Histogram},'' {\em Microscopy and Microanalysis}, vol.~6, pp.~437--444, 08
  2002.

\end{thebibliography}


\begin{thebibliography}{1}

\bibitem{manolatou1999ieee_addDropCMT}
C.~Manolatou, M.~Khan, S.~Fan, P.~Villeneuve, H.~Haus, and J.~Joannopoulos,
  ``Coupling of modes analysis of resonant channel add-drop filters,'' {\em
  IEEE Journal of Quantum Electronics}, vol.~35, no.~9, pp.~1322--1331, 1999.

\bibitem{li2016lasRev_backscattering}
A.~Li, T.~Van~Vaerenbergh, P.~De~Heyn, P.~Bienstman, and W.~Bogaerts,
  ``Backscattering in silicon microring resonators: a quantitative analysis,''
  {\em Laser \& Photonics Reviews}, vol.~10, no.~3, pp.~420--431, 2016.

\bibitem{kresse1996prb_dft}
G.~Kresse and J.~Furthm\"uller, ``Efficient iterative schemes for ab initio
  total-energy calculations using a plane-wave basis set,'' {\em Phys. Rev. B},
  vol.~54, pp.~11169--11186, Oct 1996.

\bibitem{blochl1994prb_dft}
P.~E. Bl\"ochl, ``Projector augmented-wave method,'' {\em Phys. Rev. B},
  vol.~50, pp.~17953--17979, Dec 1994.

\bibitem{perdew2008prl_dft}
J.~P. Perdew, A.~Ruzsinszky, G.~I. Csonka, O.~A. Vydrov, G.~E. Scuseria, L.~A.
  Constantin, X.~Zhou, and K.~Burke, ``Restoring the density-gradient expansion
  for exchange in solids and surfaces,'' {\em Phys. Rev. Lett.}, vol.~100,
  p.~136406, Apr 2008.

\end{thebibliography}
\end{document}


\title{Supplementary information for ``Silicon-lattice-matched boron-doped gallium phosphide: A scalable acousto-optic platform"}

\author{Nicholas S. Yama}
\thanks{nsyama@uw.edu; These two authors contributed equally}
\affiliation{University of Washington, Electrical and Computer Engineering Department, Seattle, WA, 98105, USA}%

\author{I-Tung Chen}
\thanks{These two authors contributed equally}
\affiliation{University of Washington, Electrical and Computer Engineering Department, Seattle, WA, 98105, USA}%

\author{Srivatsa Chakravarthi}
\affiliation{University of Washington, Physics Department, Seattle, WA, 98105, USA}%

\author{Bingzhao Li}
\affiliation{University of Washington, Electrical and Computer Engineering Department, Seattle, WA, 98105, USA}%

\author{Christian Pederson}
\affiliation{University of Washington, Physics Department, Seattle, WA, 98105, USA}%

\author{Bethany E. Matthews}
\affiliation{Energy and Environment Directorate, Pacific Northwest National Laboratory, Richland, Washington 99352, USA}%

\author{Steven R. Spurgeon}
\affiliation{University of Washington, Physics Department, Seattle, WA, 98105, USA}%
\affiliation{Energy and Environment Directorate, Pacific Northwest National Laboratory, Richland, Washington 99352, USA}%

\author{Daniel E. Perea}
\affiliation{Earth and Biological Sciences Directorate, Environmental Molecular Sciences Laboratory, Pacific Northwest National Laboratory, Richland, Washington 99352, USA}%

\author{Mark G. Wirth}
\affiliation{Earth and Biological Sciences Directorate, Environmental Molecular Sciences Laboratory, Pacific Northwest National Laboratory, Richland, Washington 99352, USA}%

\author{Peter V. Sushko}
\affiliation{Physical and Computational Sciences Directorate, Pacific Northwest National Laboratory, Richland, Washington 99352, USA}%

\author{Mo Li}
\affiliation{University of Washington, Electrical and Computer Engineering Department, Seattle, WA, 98105, USA}
\affiliation{University of Washington, Physics Department, Seattle, WA, 98105, USA}%

\author{Kai-Mei C. Fu}
\affiliation{University of Washington, Electrical and Computer Engineering Department, Seattle, WA, 98105, USA}
\affiliation{University of Washington, Physics Department, Seattle, WA, 98105, USA}
\affiliation{Physical Sciences Division, Pacific Northwest National Laboratory, Richland, Washington 99352, USA}%

\date{\today}

\maketitle            

\clearpage
\section{Device fabrication}
\subsection{Membrane transfer for hybrid devices}
The commercial BGaP-on-Si wafer was covered in a protective photoresist layer (AZ1512, MicroChemicals) and diced into 1\,cm$\times$1\,cm chips.
For the chips used in the hybrid devices, the silicon backside was mechanically thinned to 75\,\textmu m.
The protective resist was stripped in solvent and a new layer of AZ1512 was spun on.
Algined-mask photolithography (ABM) was then performed in order to define individual membranes with vias for wet transfer.
Individual membranes are 1.5\,mm$\times$1.5\,mm and have rectangular vias defined at 250\,\textmu m intervals.
After developing the resist, the exposed BGaP is etched in an Ar:Cl$_2$:N$_2$ inductively coupled plasma reactive ion etch (ICP-RIE) to transfer the membrane pattern into the BGaP layer.
The chip is then cleaved between the membranes for individual use.
A single-membrane chip is then flipped onto a sapphire substrate (silicon side facing up) and the sample is exposed to a XeF$_2$ vapor etch (SPTS) to remove the remaining silicon.
The BGaP membrane adheres to the sapphire substrate which is then submerged in water, releasing the membrane onto the surface (the resist is hydrophobic and orients upward).
The membrane can then be picked up on the target substrate (SiN$_x$ in this work) and allowed to dry.
After drying the adhesion to the substrate increases and the remaining photoresist can be stripped in solvent.

\subsection{Photonic device fabrication}
Starting with either a BGaP-on-Si or transferred BGaP-on-substrate chip, the sample is cleaned in a solvent rinse and allowed to dehydrate on a hotplate at 150$^\circ$C.
Electron-beam lithography (EBL) resist, hydrogen silsesquioxane (HSQ), is spun on to around 100\,nm thick and soft baked at 80$^\circ$C.
The photonic structures are then patterned in EBL and the HSQ is developed in a 25\% tetramethylammonium hydroxide (TMAH) bath.
The device pattern is transfered to the BGaP layer by ICP-RIE either completely etching through the membrane (as in the hybrid devices) or partially etching the membrane.
If the devices need to be suspended then AZ1512 photoresist is spun on and etch vias are patterned using an aligned direct write (Heidelberg DWL66 plus).
The resist is developed and the etch vias are defined in ICP-RIE.
At this stage XeF$_2$ vapor etch is used to undercut the silicon substrate for the suspended devices.

\subsection{Suspended AO device fabrication}
The BGaP photonic waveguides and grating couplers are patterned with EBL using PMMA resist. The pattern is transferred into BGaP by partial etching via ICP-RIE using chlorine-based chemistry. The 290\,nm thick ZnO film is deposited using an RF magnetron sputtering system and lifted-off in a sonicated acetone bath. The IDT pattern is written using the EBL and lifted-off after depositing 220\,nm aluminum film using an electron-beam evaporator. The releasing vias are patterned with photolithography and etched in ICP. Finally, XeF$_2$ etching of silicon is used for suspending the BGaP layer.

\clearpage
\begin{figure*}[t]
    \centering
    \includegraphics[width=\textwidth]{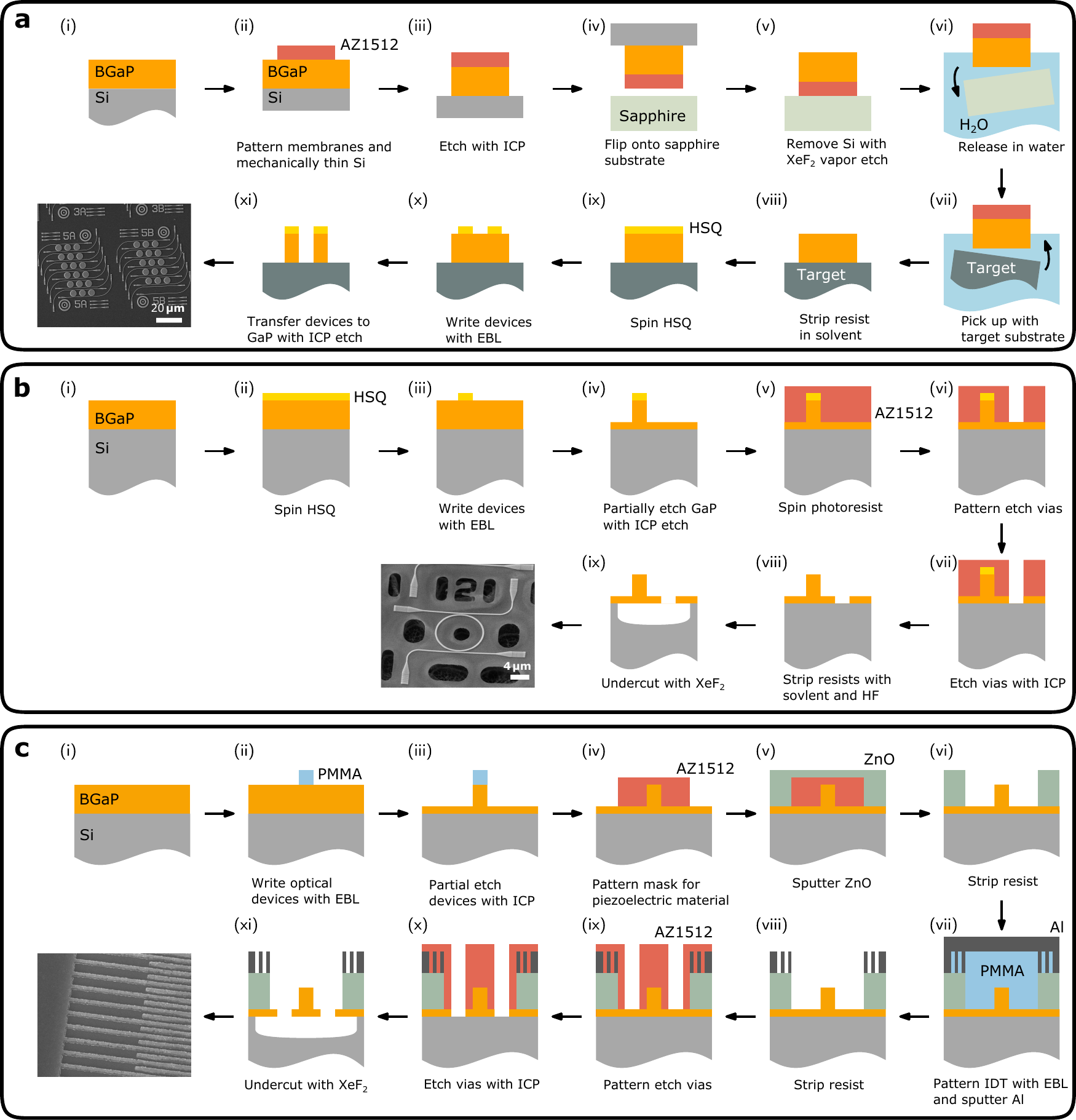}
    \caption{\textbf{Fabrication process} Schematic representations and end results of the fabrication processes for \textbf{(a)} hybrid devices, \textbf{(b)} suspended photonics, and \textbf{(c)} supsended AO devices.
    }
    \label{fig:fabrication_process}
\end{figure*}

\clearpage
\begin{figure*}[h]
    \centering
    \includegraphics[width=\textwidth]{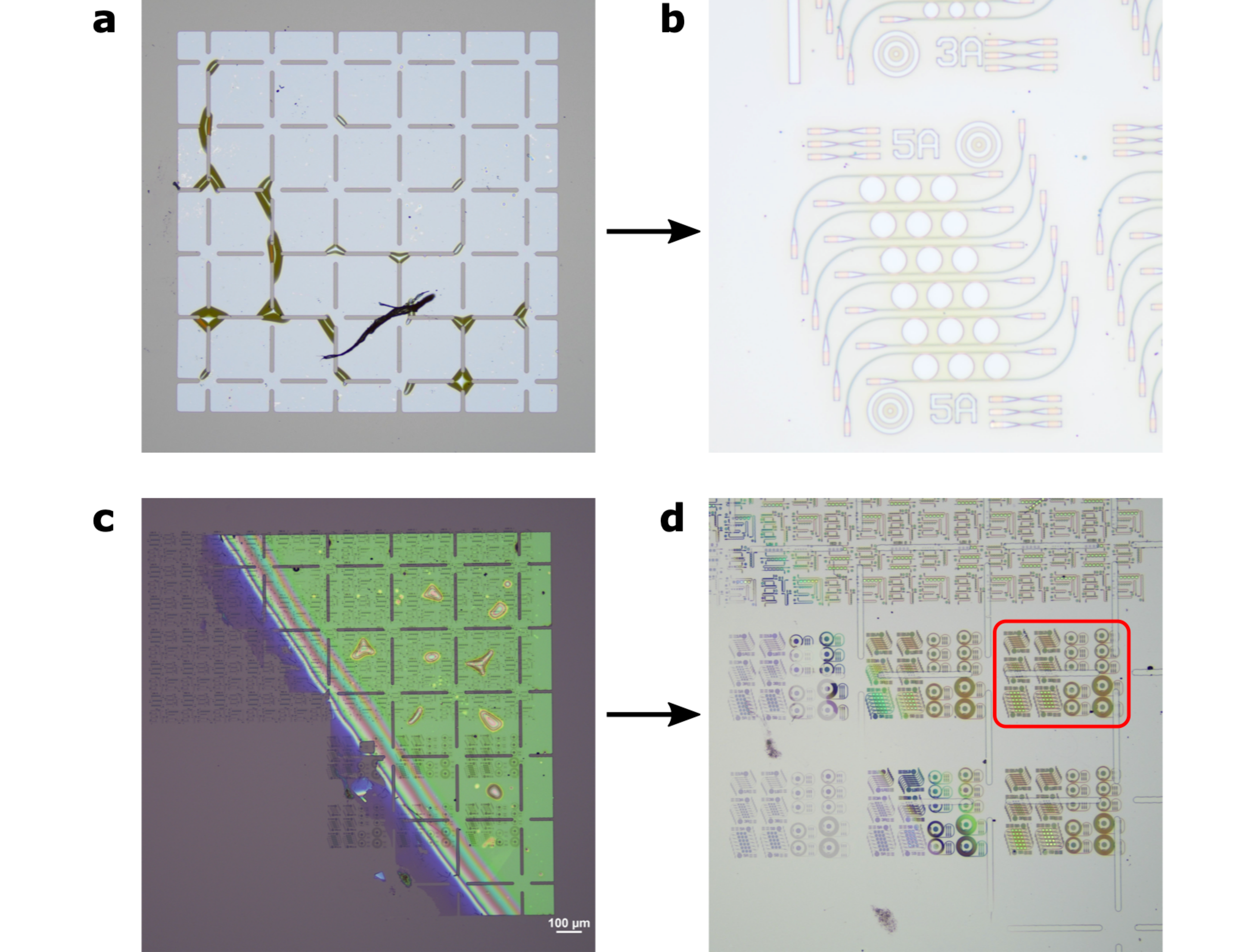}
    \caption{\textbf{Membrane transfer optical images.}
    Optical images of the hybrid BGaP (\textbf{a}, \textbf{b}) and the un-doped MBE GaP (\textbf{c}, \textbf{d}) devices.
    \textbf{(a)} Transferred BGaP membrane for the devices used in the paper (approximately 1.5\,mm on both edges, vias are spaced at 250\,\textmu m).
    Several bubbles formed on the bottom-left side of the membrane.
    The large particle on the bottom is suspected to be some of the photoresist which did not immediately dissolve.
    The remaining top-right corner is well-adhered and used to pattern devices.
    \textbf{(b)} One of the final BGaP 5-\textmu m disk devices.
    \textbf{(c)} The transferred un-doped GaP membrane after device patterning (prior to etching).
    The membrane is partial as it is from the edge of the wafer.
    \textbf{(d)} Final devices after etching.
    The highlighted devices are the high-$Q$ optical devices.
    In both cases, the design of the 5-\textmu m disks are identical.
    }
    \label{fig:fabrication_transfer}
\end{figure*}

\clearpage
\begin{figure*}[h]
    \centering
    \includegraphics[width=\textwidth]{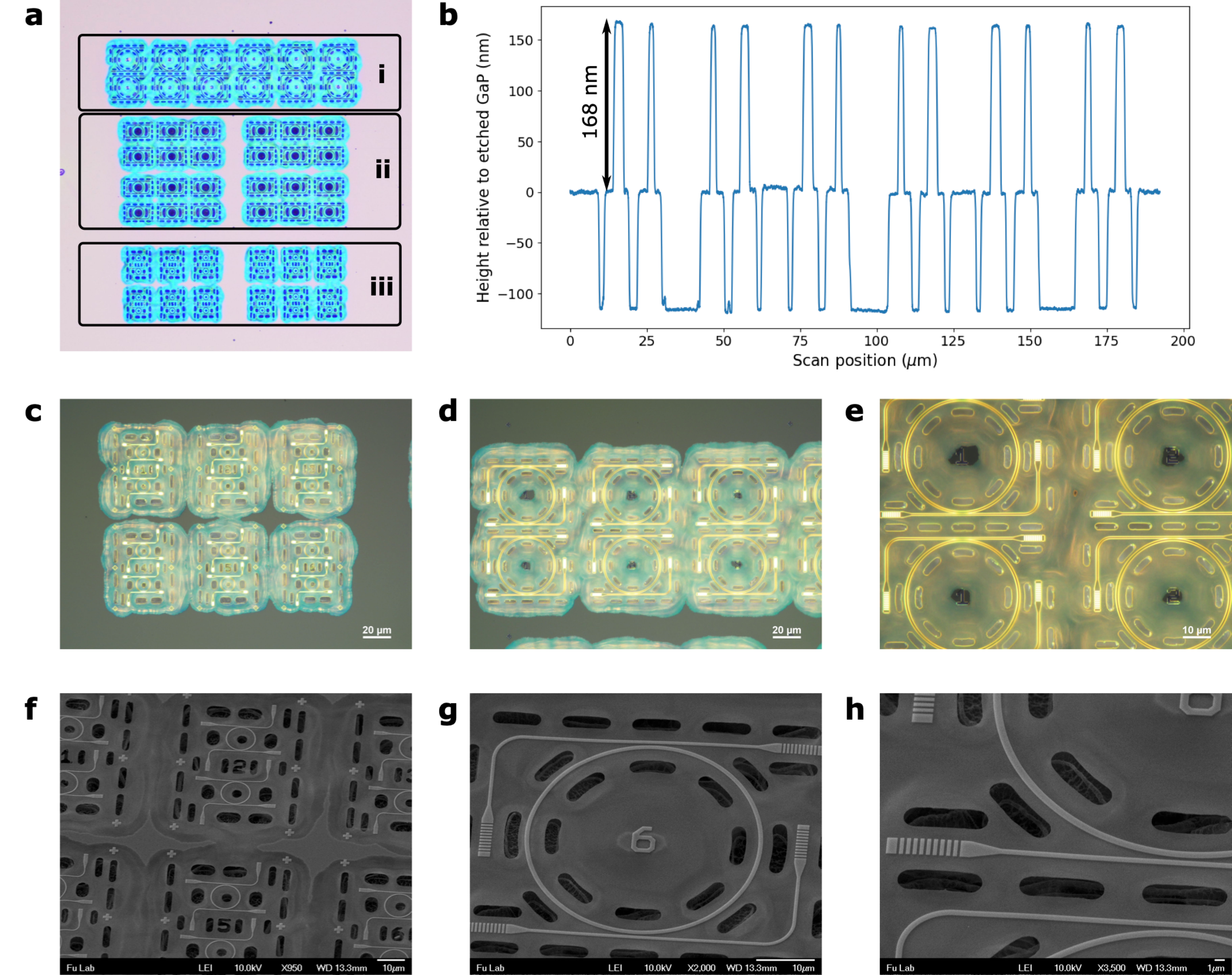}
    \caption{\textbf{Suspended devices.}
    \textbf{(a)} Real-color optical microscope image of the measured device set after fabrication.
    Encircled device sets are (i) 40-\textmu m ring telecom devices, (ii) 20-\textmu m ring telecom devices, and (iii) 10-\textmu m ring visible devices.
    The light blue color is real and corresponds to the undercut area.
    \textbf{(b)} Profilometer measurement (Bruker DektakXT) of the devices prior to undercutting the substrate with XeF$_2$.
    The height is zeroed to partially etched GaP height giving the partial etch depth to be 168\,nm.
    Likewise, the etch via depth is measured to be around 110\,nm which is consistent with the approximately 268\,nm thick membrane and some slight etching of the silicon substrate by the ICP-RIE.
    Optical microscope images of \textbf{(c)} visible devices, and \textbf{(d, e)} 40-\textmu m ring telecom devices using cross-polarized bright-field imaging.
    The membrane appears transparent and the undercut silicon surface can be seen.
    SEM images of the \textbf{(f)} 10-\textmu m ring visible devices, and \textbf{(g,h)} 40-\textmu m ring telecom devices.
    }
    \label{fig:fabrication_suspended}
\end{figure*}

\clearpage
\section{Device measurement}
\subsection{Photonic device transmission measurements}
Samples were loaded into custom-built confocal microscope systems and transmission measurements were performed using a supercontinuum laser (Fianium), scanning diode laser (Sacher), and/or scanning telecom laser (HP) sent through an erbium-doped fiber amplifier (Thorlabs).
The transmission signal was filtered through a combination of spatial filtering and cross-polarized optics.
The resulting light could be collected on a spectrometer (Princeton instruments) or avalanche-photodiodes (Excelitas, ThorLabs).


\clearpage
\section{Material characterization}
\subsection{Atomic-force microscopy}
The surface roughness of the material was measured using a Bruker Icon atomic force microscope in tapping mode under ambient conditions.

\subsection{Variable-angle spectroscopic ellipsometry}
Variable angle spectroscopic ellipsometry (J. A. Woollam RC2) is performed over GaP transparency window (500\,nm--2500\,nm) on an unused BGaP-on-Si chip to determine the refractive index and (measurable) optical absorption.
Assuming layer thicknesses of 41\,nm and 37\,nm for the top and bottom GaP layers, and using known GaP refractive index data, the BGaP layer thickness is fitted against a Gaussian oscilaltor model.
Importantly, the backside of the silicon substrate (750\,\textmu m) is polished and must be accounted for in the model.
The fitting is performed in the CompleteEASE software (J. A. Wollam) for measurements at 55$^\circ$, 65$^\circ$, and 75$^\circ$ and wavelength range between 500--2500\,nm.

\begin{figure*}[h]
    \centering
    \includegraphics[width=0.75\textwidth]{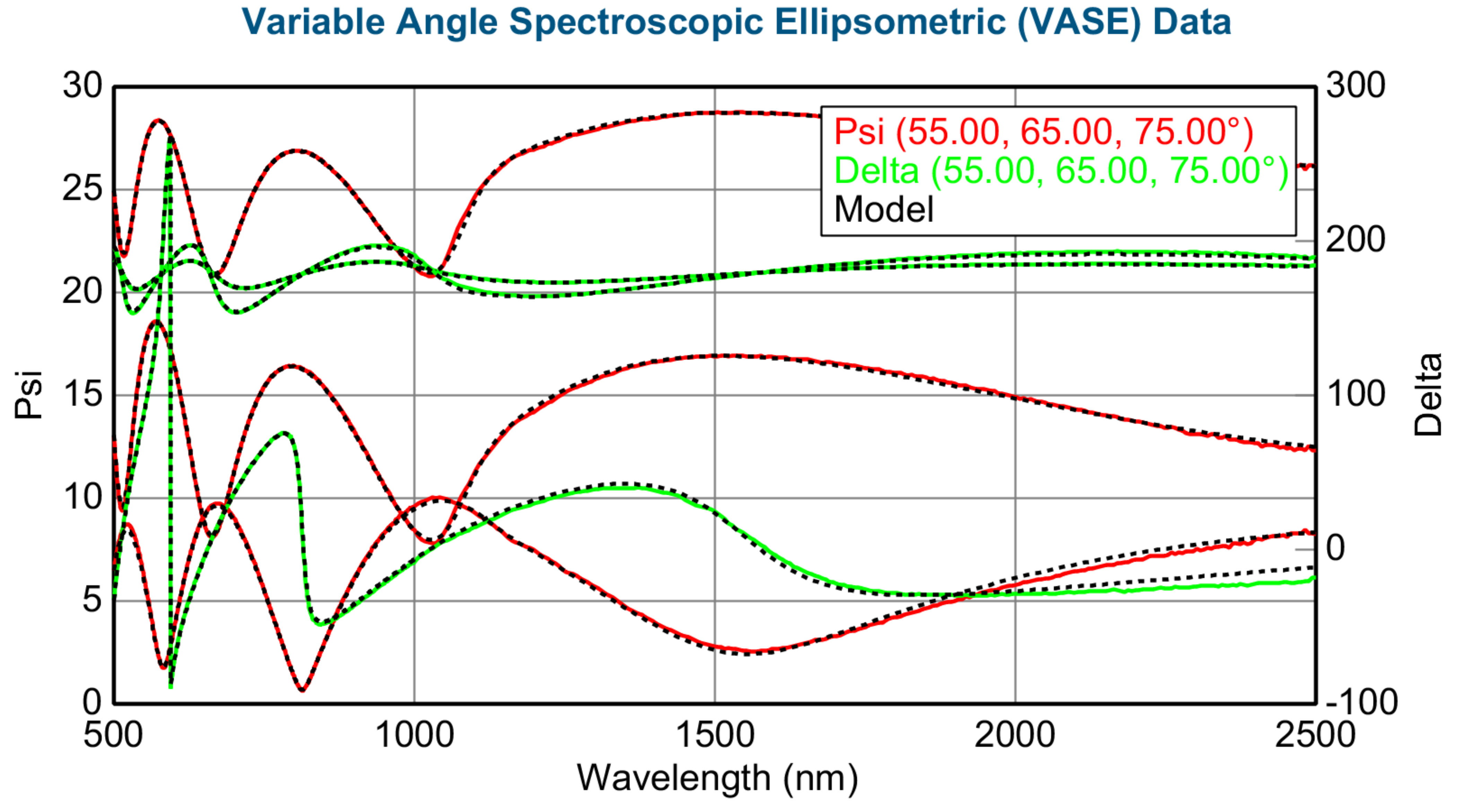}
    \caption{\textbf{VASE fit.}
    Data and fit for the VASE ellipsometry data obtained from CompleteEASE.
    }
    \label{fig:vase_fit}
\end{figure*}

\subsection{Scanning Transmission Electron Microscopy}
Cross-sectional TEM samples were prepared using a FEI Helios NanoLab DualBeam Focused Ion Beam (FIB) microscope and a standard lift out procedure. Images of the as-grown and post-irradiated samples were collected using a probe-corrected JEOL GrandARM-300F STEM operating at 300 kV with a semi-convergence angle of 27.5 mrad and a collection angle of 82--186 mrad (HAADF) and 48--150 (MAADF). Additional drift-corrected high-resolution HAADF images were collected using a probe-corrected Thermo Fisher Scientific Themis Z microscope operating at 300 kV, with a convergence semiangle of 25.2 mrad and an approximate collection angle range of 65 to 200 mrad.


\clearpage
\section{Temporal coupled-mode theory model for the add-drop resonator} \label{sec:cmt}
We model the disk/ring resonator devices using temporal coupled-mode theory (CMT).
This section derives the device transfer functions from the coupled-mode equations as used in the main text.

\begin{figure*}[h]
    \centering
    \includegraphics[width=\textwidth]{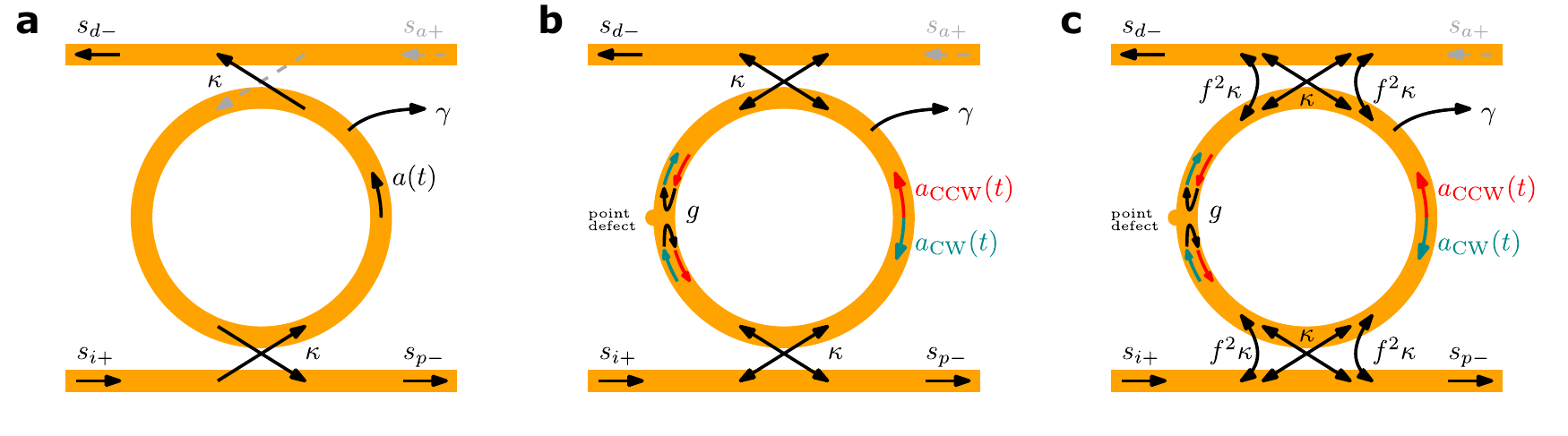}
    \caption{\textbf{Coupled-mode theory model.}
    Schematic representations of the coupling between the waveguide and cavity mode(s) for
    \textbf{(a)} ideal, \textbf{(b)} symmetrically split, and \textbf{(c)} asymmetrically split cases.
    }
    \label{fig:cmt_model}
\end{figure*}

We assume throughout that the resonance linewidths are significantly smaller than the free spectral range.
Additionally we consider a coupling geometry corresponding to the four-port add-drop resonator configuration as shown in Fig.~\ref{fig:cmt_model}.
The cavity field terms $a_j(t)$ give the amplitude and (temporally varying) phase of the $j$-th propagating cavity resonance, normalized such that $|a_j|^2$ gives the energy in the cavity.
Likewise, the waveguide field terms $s_{k\pm}$ give the incoming/outgoing waveguide field amplitudes, normalized such that $|s_{k\pm}|^2$ gives the power in the corresponding mode.
It is assumed that the only incoming mode is at the input port $s_{i+}$ and that there is no reflection near the coupling region.

\subsection{Ideal case}
We first consider the idealized case shown in Fig.~\ref{fig:cmt_model}a in which a single counter-clockwise propagating mode $a(t)$ at frequency $\omega_0$ with intrinsic decay rate $\gamma$ is coupled to each waveguide with a rate $\kappa$.
Time evolution of the mode under driving $s_{i+}$ is described by the differential equation
\begin{equation} \label{eq:single_mode_eqn}
    \dv{a}{t} = \left(-i\omega_0 + \frac{\Gamma}{2}\right) a + \sqrt{\kappa} e^{i\theta} s_{1+},
\end{equation}
where $\Gamma = \gamma + 2\kappa$ is the total \textit{energy} decay rate. 
Note the factor of $2\kappa$ corresponds to the fact that there are two symmetrical waveguides for coupling.

The outgoing fields at the drop ($s_{d-}$) and pass ($s_{p-}$) ports are related to the add port ($s_{a+}$), input field ($s_{i+}$), and cavity field $a$ through the linear system of equations \cite{manolatou1999ieee_addDropCMT}
\begin{align}
    s_{p-} &= e^{i\phi} (s_{i+} - \sqrt{\kappa} e^{-i\theta} a) ,\\
    s_{d-} &= e^{i\phi} (s_{a+} - \sqrt{\kappa} e^{-i\theta} a) ,
\end{align}
where $\phi$ corresponds to the phase winding incurred between the port reference planes.
Here and throughout we assume that $s_{a+} = 0$.
These equations in combination with equation~\eqref{eq:single_mode_eqn} constitute the coupled-mode equations for a single-mode resonator in the add-drop configuration.

Since we are interested in the steady-state solution for a periodic driving at frequency $\omega$, we can solve the equations by Fourier transform taking $a(t) \to A(\omega)$, and likewise $s(t) \to S(\omega)$ for the waveguide fields.
To determine the response functions for the pass $T_{p}$ and drop $T_{d}$ ports we solve the coupled-mode equations as
\begin{align}
    T_{p} = \abs{\frac{S_{p-}}{S_{i+}}}^2 = \frac{\Delta^2 + (\gamma/2)^2}{\Delta^2 + (\gamma/2 + \kappa)^2} , \\
    T_{d} = \abs{\frac{S_{d-}}{S_{i+}}}^2 = \frac{\kappa^2}{\Delta^2 + (\gamma/2 + \kappa)^2},
\end{align}
where $\Delta\equiv\omega_0 - \omega$ is the detuning.
These correspond to a Lorentzian dip/peak at the resonance frequency $\omega_0$ with full-width-half-max (FWHM) of $\Gamma = \gamma + 2\kappa$.

The total quality factor $Q$ of the resonator at this resonance is defined as the ratio $Q \equiv \omega_0 / \Gamma$.
We can further distinguish the intrisic quality factor $Q_i \equiv \omega_0 / \gamma$ and coupling quality factor $Q_c \equiv \omega_0/\kappa$ which distinguish between intrinsic and coupling losses respectively.
The quality factors are related by the inverse relationship $1/Q = (1/Q_i) + (2/Q_c)$.

The depth of the resonance in the pass port is given by
\begin{equation}
    T_p(\Delta = 0) = \frac{(\gamma/2)^2}{(\gamma/2 + \kappa)^2} = \left(\frac{Q_T}{Q_i}\right)^2.
\end{equation}
This allows for a straightforward determination of the intrinsic quality factor of the resonator from a single measurement of the through port provided the resonance depth and total $Q$ can be determined accurately.

\subsection{Symmetric mode splitting}
We now consider the case of symmetric splitting as schematically shown in Fig.~\ref{fig:cmt_model}b.
The ring resonator structure supports two degenerate, counter-propagating modes at a given frequency; one clockwise propagating (CW) and one counter-clockwise (CCW).
In the previous section we considered the CCW mode, which, under ideal conditions, would remain independent of the CW mode.
However, imperfections in the fabricated resonator can introduce defects which cause \textit{back-scatter} thus coupling the modes at some rate $g$ and splitting the resonance profile.
In the figure, a single point defect scatterer is shown, however the observed coupling will generally be the result of many such defects distributed over the entire resonator.

The coupled-mode equations are modified to include the CW-propagating mode which, by symmetry, has an identical resonance frequency $\omega_0$ and loss rate $\Gamma$.
Letting $a_{\mathrm{CCW}}$ ($a_{\mathrm{CW}}$) be the CCW (CW) modes respectively, we have
\begin{align}
    \dv{a_{\mathrm{CCW}}}{t} &= \left(-i\omega_0 + \frac{\Gamma}{2}\right) a_{\mathrm{CCW}} - i g a_{\mathrm{CW}} + \sqrt{\kappa} e^{i\theta} s_{i+}, \\
    \dv{a_{\mathrm{CW}}}{t} &= \left(-i\omega_0 + \frac{\Gamma}{2}\right) a_{\mathrm{CW}} - i g^* a_{\mathrm{CCW}} , \\
    s_{p-} &= e^{i\phi} (s_{i+} - \sqrt{\kappa} e^{-i\theta} a_{\mathrm{CCW}}) ,\\
    s_{d-} &= e^{i\phi} (s_{a+} - \sqrt{\kappa} e^{-i\theta} a_{\mathrm{CCW}}) ,
\end{align}
where the coupling is introduced through the terms containing $g$ which corresponds to the coupling rate.
This form implicitly assumes that the back-scattering process is energy conserving which is a good approximation in most cases \cite{li2016lasRev_backscattering}. 
The Fourier transform method can be applied again to determine
\begin{align} \label{eq:ideal_case_pass}
    T_p &= \abs{1 - \frac{\kappa (i\Delta + \Gamma)}{(i\Delta + \Gamma)^2 + |g|^2}}^2 , \\
    T_d &= \abs{\frac{\kappa (i\Delta + \Gamma)}{(i\Delta + \Gamma)^2 + |g|^2}}^2.
\end{align}

\subsection{Asymmetric mode splitting}
The back-scattering process described in the previous section produces only symmetrically split modes.
However, in practice, modes are often asymmetrically split.
This is attributed to additional \textit{back-coupling} mechanisms \cite{li2016lasRev_backscattering} in which the resonator field is partially reflected at the waveguide-ring coupling region due to the change in effective index.
Some amount of the reflected light couples into the counter-propagating waveguide mode.
This is shown schematically in Fig.~\ref{fig:cmt_model}c.

To model this the coupled-mode equations are modified to take the form
\begin{align}
    \dv{a_{\mathrm{CCW}}}{t} &= \left(-i\omega_0 + \frac{\Gamma}{2}\right) a_{\mathrm{CCW}} - i g a_{\mathrm{CW}} + \sqrt{\kappa} e^{i\theta} s_{i+}, \\
    \dv{a_{\mathrm{CW}}}{t} &= \left(-i\omega_0 + \frac{\Gamma}{2}\right) a_{\mathrm{CW}} - i g^* a_{\mathrm{CCW}}  + f \sqrt{\kappa} e^{i\theta} s_{i+}, \\
    \label{eq:full_split_thru}
    s_{p-} &= e^{i\phi} (s_{i+} - \sqrt{\kappa} e^{-i\theta} a_{\mathrm{CCW}} - f^* \sqrt{\kappa} e^{-i\theta} a_{\mathrm{CW}}) ,\\
    \label{eq:full_split_drop}
    s_{d-} &= e^{i\phi} (s_{a+} - \sqrt{\kappa} e^{-i\theta} a_{\mathrm{CCW}} - f^* \sqrt{\kappa} e^{-i\theta} a_{\mathrm{CW}}) ,
\end{align}
where we have introduced the back-coupling coefficient $f \in \mathbb{C}$ from \cite{li2016lasRev_backscattering} with the modification of a complex conjugate in the output coupling terms.
The back-coupling coefficient parameterizes the relative amplitude and phase of the back-coupling rate with respect to the complex forward-coupling rate $\kappa e^{2i\theta}$.
Because the back-coupling is another loss channel, the total loss rate $\Gamma$ must be modified to include its contribution giving $\Gamma = \gamma + 2 (1 + |f|^2) \kappa$.

Taking the Fourier transform as before allows for the determination of the system response.
The solutions for $A_{\mathrm{CCW}}$ and $A_{\mathrm{CW}}$ are found to be
\begin{align}
    A_{\mathrm{CCW}} &= \frac{-igf\sqrt{\kappa}e^{i\theta} + (i\Delta + \Gamma) }{(i\Delta + \Gamma)^2 + |g|^2}, \\
    A_{\mathrm{CW}} &= -\frac{|g|^2 f\sqrt{\kappa}e^{i\theta} }{(i\Delta + \Gamma) \big[(i\Delta + \Gamma)^2 + |g|^2\big]} 
        - \frac{ig^*\sqrt{\kappa}e^{i\theta}}{(i\Delta + \Gamma)^2 + |g|^2}
        + \frac{f\sqrt{\kappa}e^{i\theta}}{(i\Delta + \Gamma)}.
\end{align}
Substituting this into Eqs.~\eqref{eq:full_split_thru} and \eqref{eq:full_split_drop} and simplifying terms gives the following system response functions for the pass and drop ports
\begin{align}
    T_p &= \abs{1 + \frac{2i\kappa|g||f|\cos\beta - \kappa(1+|f|^2) (i\Delta + \Gamma)}{(i\Delta + \Gamma)^2 + |g|^2}}^2, \\
    T_d &= \abs{\frac{2i\kappa|g||f|\cos\beta - \kappa(1+|f|^2) (i\Delta + \Gamma)}{(i\Delta + \Gamma)^2 + |g|^2}}^2,
\end{align}
where we have parameterized the splitting in terms of $(|g|, |f|, \beta)$ where $\beta = \arg(gf)$.

The response functions can be written in the more suggestive form of
\begin{align}
    T_p &= |1 + X|^2, \\
    T_d &= |X|^2,
\end{align}
where
\begin{equation} \label{eq:asymmetric_exact}
    X \equiv -\frac{\kappa}{2} \left(
        \frac{(|f|+e^{i\beta})(|f|+e^{-i\beta})}{i(\Delta + |g|) + \Gamma} + 
        \frac{(|f|-e^{i\beta})(|f|-e^{-i\beta})}{i(\Delta - |g|) + \Gamma}
    \right).
\end{equation}
This shows that the parameters have simple physical interpretations with $2|g|$ defining the splitting and $(|f|,\beta)$ giving the amplitude and direction of the asymmetry.
This reproduces the form of \cite{li2016lasRev_backscattering} with the addition of tracking the relative phase between the two coupling processes.

\subsubsection{Weak back-coupling approximation}
Although this model can be applied to a wide range of observed resonances, fitting can be challenging in practice.
In particular, the number of parameters and their relative co-linearity prevents high-confidence fitting from a single measured response function (typically the pass).
Instead transmission at the pass, drop, and add ports need to be considered simultaneously.
This complicates the measurement significantly as it not only requires measurements of the weaker drop/add signals but also necessitates thorough characterization of optical components other than the resonator itself (e.g. the grating couplers, waveguides, and collection optics).

Instead we make an approximation of weak back-coupling such that $|f|\ll1$.
This approximation is intuitive and is numerically demonstrated in Section~\ref{sec:optical_fitting} in which the noisy (exact model) fits closely reproduce the approximate model results while estimating $|f|<0.01$.
Enforcing this approximation allows for the quadratic terms $|f|^2$ to be dropped.
As a result we have
\begin{equation}
    X \approx -\frac{\kappa}{2} \left(
        \frac{1 + 2|f|\cos\beta}{i(\Delta + |g|) + \Gamma} + 
        \frac{1 - 2|f|\cos\beta}{i(\Delta - |g|) + \Gamma}
    \right)
\end{equation}
and $\Gamma \approx \gamma + 2 \kappa$ as before.
This reduces the number of free parameters by one and gives $|f|\cos\beta$ as a single \textit{real-valued} parameter describing the asymmetry of the power spectrum.

Thus we determine the transmission at the pass and drop ports to be given by
\begin{align} \label{eq:pass_small_asym}
    T_p &= 
    \left|1
    -\frac{\kappa}{2} \left(
        \frac{1 + 2|f|\cos\beta}{i(\Delta + |g|) + \Gamma} + 
        \frac{1 - 2|f|\cos\beta}{i(\Delta - |g|) + \Gamma}
    \right)
    \right|^2 , \\
    T_d &= 
    \left|
    \frac{\kappa}{2} \left(
        \frac{1 + 2|f|\cos\beta}{i(\Delta + |g|) + \Gamma} + 
        \frac{1 - 2|f|\cos\beta}{i(\Delta - |g|) + \Gamma}
    \right)
    \right|^2.
\end{align}
This form of the pass transmission is used for the fitting to determine the parameters of the high-$Q$ telecom resonances of the main text.

\clearpage
\section{Fitting of optical data} \label{sec:optical_fitting}
The CMT analysis discussed in Section \ref{sec:cmt} provides analytical functions for the resonance lineshape.
This section covers specific details of the fitting procedure relevant to reproduce the results.
In all cases the data is fit to the model as determined by the a nonlinear-least-squares optimization algorithm (Scipy).
Additional fits and data not shown in the main text are also included for completeness.

\subsection{Drop port transmission}
In the case of drop port spectra, the absolute transmission value cannot typically be ascertained without careful (and often noisy) calibration of the grating coupler and waveguide transmission spectra.
Fortunately, for the purposes of this work, we need only determine the linewidth of the resonance and so the lineshape can be fitted against a generic Lorentzian 
\begin{equation}
    T_d^{(1)}(x) = a \frac{(\Gamma/2)^2}{(x-x_0)^2 + (\Gamma/2)^2} + p(x-x_0)
\end{equation}
where $p(x) = p_0 + p_1 x + p_2 x^2$ is a quadratic background centered at the resonance center.
The parameter $a$ corresponds to the amplitude, $\Gamma$ is the full-width-half-max of the resonance, and $x_0$ corresponds to the center position.

The above functional form assumes that the background spectra (e.g. transfer function of the waveguide, scattered input, dark counts) varies slowly over the resonance linewidth which is typically confirmed by the symmetrical lineshape of most drop port resonances.
Strongly asymmetrical lineshapes correspond to interference with additional modes, such as with Fano resonances, and can have Lorentzian-fitted linewidths which do not correspond directly to the resonance loss.
As such, these modes were excluded from this work where possible.

For the high-$Q$ intrinsic GaP devices, the laser scaning measurements reveal split modes.
For these resonances a simplified version of Eq.~\eqref{eq:asymmetric_exact} was used for the fitting procedure given as
\begin{equation}
    T_d^{(2)} =  \left|
        \frac{a+\delta a}{i((x_0 + \delta x_0) - x) + (\Gamma + \delta\Gamma)} + 
        \frac{a-\delta a}{i((x_0 - \delta x_0) - x) + (\Gamma - \delta\Gamma)}
    \right|^2 + p(x-x_0).
\end{equation}
where the $\delta a$, $\delta x_0$, and $\delta\Gamma$ terms are small shifts to the amplitude, peak centers, and linewidths respectively.
The $Q$ is then extracted assuming that these perturbations were symmetric about the original mode which is consistent with the CMT model.

\subsection{Pass port transmission}
The pass port transmission signal is advantagoues in that the normalization can be determined directly from the spectrum provided the noise and ambient background is well characterized.
Pass port transmission spectra for the low-$Q$ resonances were fit with the ideal CMT model given in Eq.~\eqref{eq:ideal_case_pass} with the modification of a multiplicative background to account for transmission through the grating and waveguide, given as
\begin{equation}
    T_p^{(1)}(\omega) =  \frac{(\omega_0 - \omega)^2 + (\gamma/2)^2}{(\omega_0 - \omega)^2 + (\gamma/2 + \kappa)^2} \cdot p(\omega-\omega_0),
\end{equation}
where again $p(x)=p_0 + p_1 x + p_2 x^2$ is the quadratic background modulation.
In the case of split modes, the pass transmission is fit to the model given in Eq.~\eqref{eq:pass_small_asym} with form
\begin{equation}
    T_p^{(1)} = 
    \left|1
    -\frac{\kappa}{2} \left(
        \frac{1 + 2|f|\cos\beta}{i((\omega_0 - \omega) + |g|) + (\gamma + 2\kappa)} + 
        \frac{1 - 2|f|\cos\beta}{i((\omega_0 - \omega) - |g|) + (\gamma + 2\kappa)}
    \right)
    \right|^2 \cdot p(\omega - \omega_0)
\end{equation}
where $\{ \omega_0, |f|\cos\beta, \kappa,  \gamma, |g| \}$ are used as the fitting parameters (in addition to the 3 parameters for $p(x)$).

\begin{figure*}[h]
    \centering
    \includegraphics{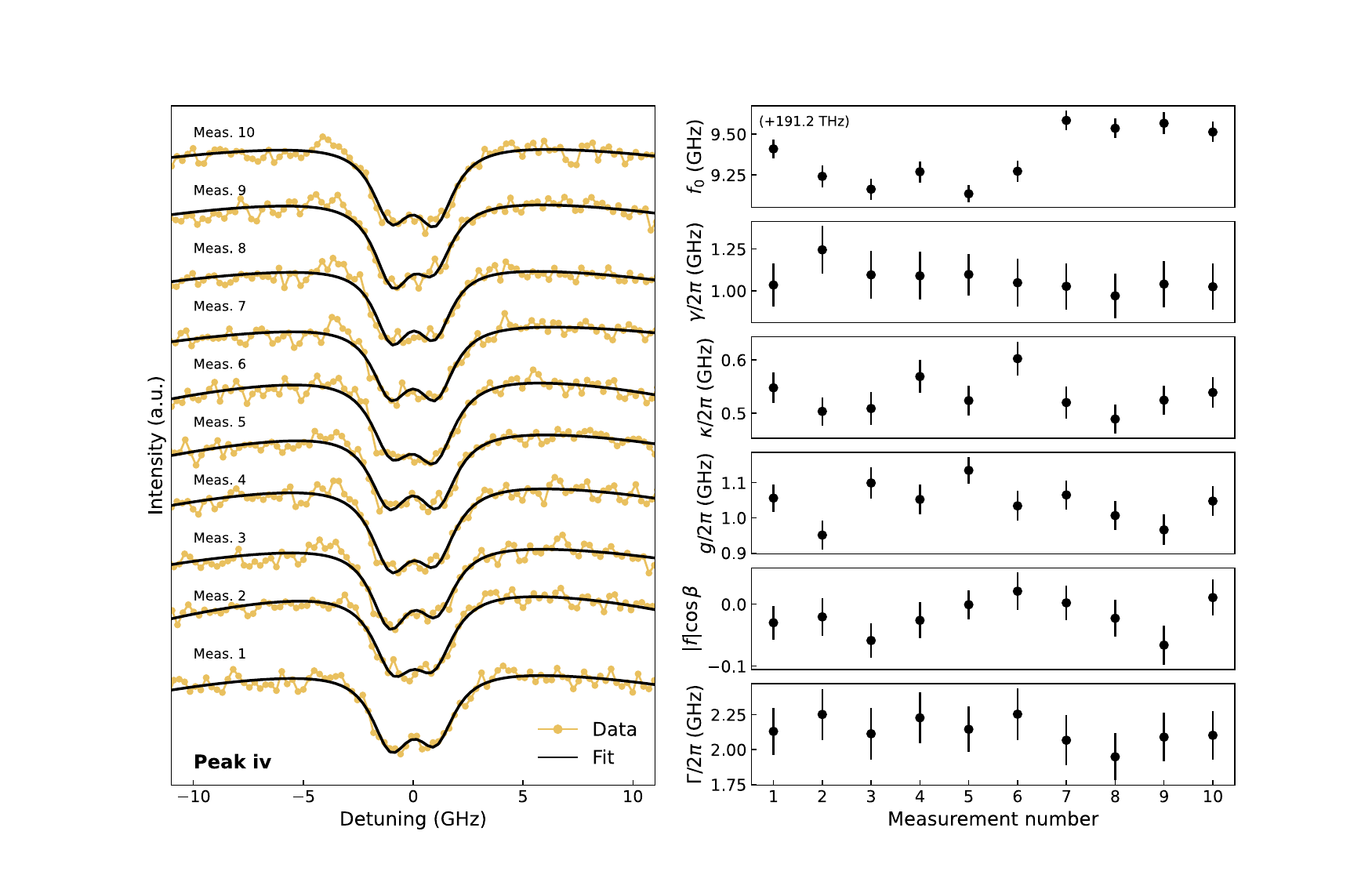}
    \caption{\textbf{High-$Q$ resonator measurements.}
    Example set of measurements for the high-$Q$ resaonce labeled as peak (vi) in the main text, Fig.~3, and corresponding fit parameters for each individual measurement.
    }
    \label{fig:fitting}
\end{figure*}

Beyond the mode splitting, another consequence of the high $Q$ factors was thermal optical bi-stability which results in asymmetric mode profiles.
This would prevent accurate estimation of the quality factor.
To combat this we utilized stepped sweeps at lower optical powers for which the bistability effect was negligible.
However, the lower powers yield reduced signal-to-noise and so multiple measurements of the same resonance were employed.
These measurements could then be fitted individually to the split-mode model and then the resulting parameters could be averaged.
An example of the multiple scans are shown in Fig.~\ref{fig:fitting} along with the corresponding fit parameters for one resonance.
Given such a set of measurements indexed by $i$, each with fit parameters $\mu_i$ and associated variance $\sigma_i^2$, the most-likely estimator for the parameter $\mu$ is given by
\begin{equation}
    \mu \equiv \frac{1}{\sum_i (1/\sigma_{i}^2)} \sum_i \frac{\mu_i}{\sigma_{i}^2}
\end{equation}
and has variance
\begin{equation}
    \sigma^2 \equiv \frac{1}{\sum_i (1/\sigma_{i}^2)}.
\end{equation}
These values are used to calculate the quality factors of the high-$Q$ telecom resonances in the main text.
The fitted parameters and associated errors as determined by this procedure for the four high-$Q$ resonances shown in the main text are listed below in Table~\ref{tab:fit_params}.

\begin{table}[h]

    \caption{\label{tab:fit_params}%
        Fitted parameters for high-$Q$ telecom resonances in main text Fig.~3.
    }
    \begin{ruledtabular}
        \begin{tabular}{ccccccc}
        \textrm{Resonance}&
        $\omega_0/2\pi$&
        $\gamma/2\pi$ (THz)&
        $\kappa/2\pi$ (GHz)&
        $g/2\pi$ (GHz)&
        $|f|\cos\beta$ (u.l.)&
        $\Gamma/2\pi$ (GHz)
        \\
        
    \colrule

    i & 
        $194.001 \pm 0.000$ &
        $0.862 \pm 0.049$ &
        $0.445 \pm 0.010$ &
        $0.596 \pm 0.016$ &
        $-0.049 \pm 0.020$ &
        $1.747 \pm 0.064$ \\

    ii & 
        $193.311 \pm 0.000$ &
        $1.061 \pm 0.049$ &
        $0.513 \pm 0.010$ &
        $0.892 \pm 0.0147$ &
        $-0.036 \pm 0.012$ &
        $2.095 \pm 0.063$ \\

    iii & 
        $192.620 \pm 0.000$ &
        $0.969 \pm 0.061$ &
        $0.613 \pm 0.013$ &
        $0.613 \pm 0.013$ &
        $-0.023 \pm 0.022$ &
        $2.196 \pm 0.079$ \\

    iv & 
        $191.929 \pm 0.000$ &
        $1.066 \pm 0.043$ &
        $0.530 \pm 0.009$ &
        $1.043 \pm 0.013$ &
        $-0.018 \pm 0.009$ &
        $2.129 \pm 0.056$ \\

        \end{tabular}
    \end{ruledtabular}

\end{table}

\subsubsection{Validity of the weak back-coupling approximation}
\begin{figure*}[h]
    \centering
    \includegraphics{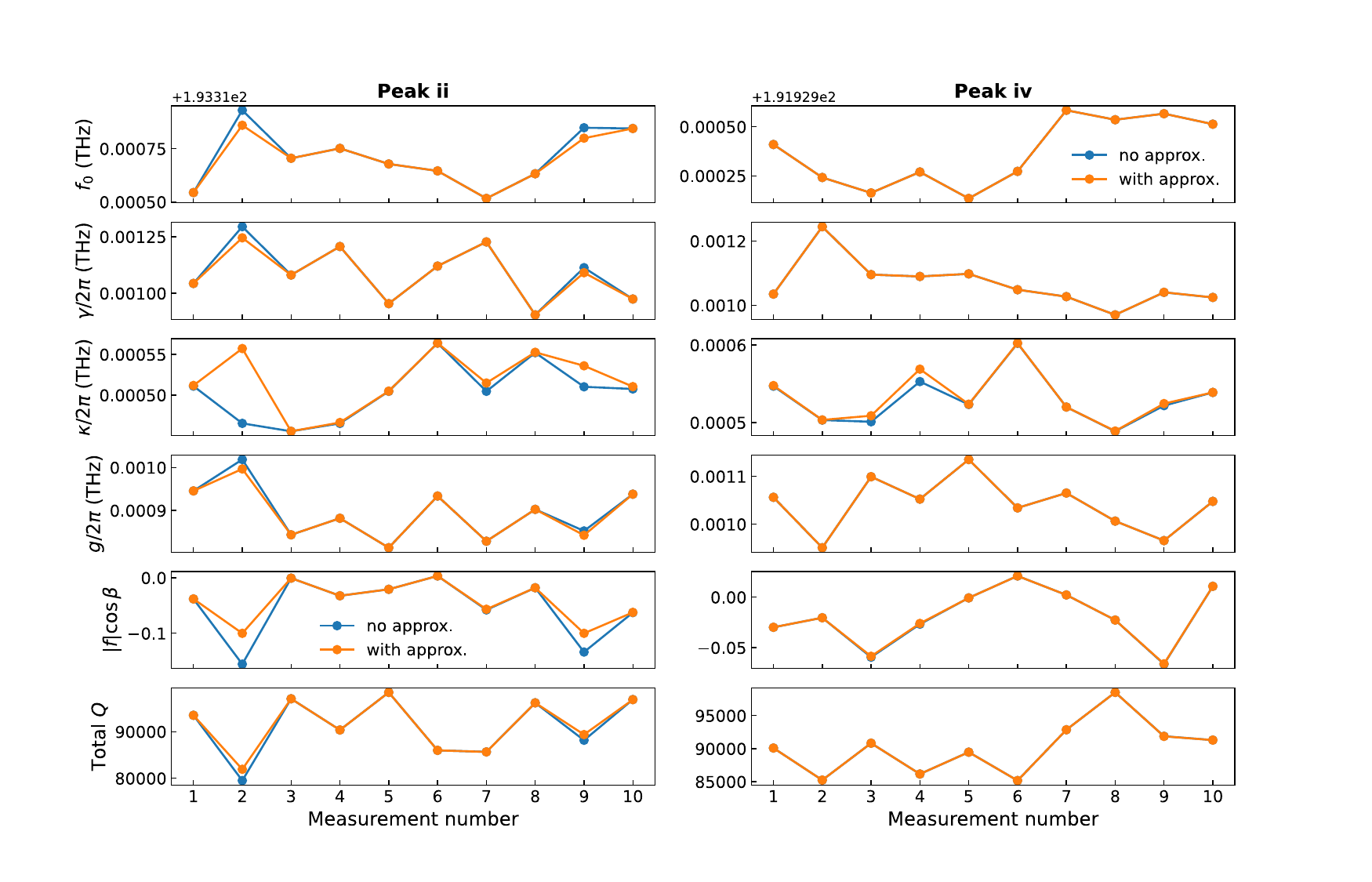}
    \caption{\textbf{Unconstrained fitting.}
    Comparison between the fitted parameters obtained by the unconstrained (blue) and approximate (orange) asymmetric split resonance models for two peaks correspondingly labeled in the main text Fig.~3.
    In nearly all cases the parameters are remarkably similar in value with and without the inclusion of the weak back-coupling approximation.
    }
    \label{fig:limit_verfication}
\end{figure*}

In the previous section, the weak back-coupling approximation was used as the full model over-fits individual pass spectra.
To validate this approximation, we fit using the full model given in Eq.~\eqref{eq:asymmetric_exact} without constraining the parameters, comparisons between the constrained and unconstrained models are shown in Fig.~\ref{fig:limit_verfication}.
In nearly all cases the parameters are remarkably similar in both methods with almost identical values.
The few cases for which the fitting results do not converge typically correspond to the model erroneously interpreting a background feature as a second peak which is not present in other measurements of the same resonance.
In all fits where such issues are not present, the resulting total and intrinsic quality factors are nearly identical, with deviations of at most a few percent.
Similarly the estimated effective back-coupling parameter $|f|\cos\beta$ similarly converges.

\begin{figure*}[h]
    \centering
    \includegraphics{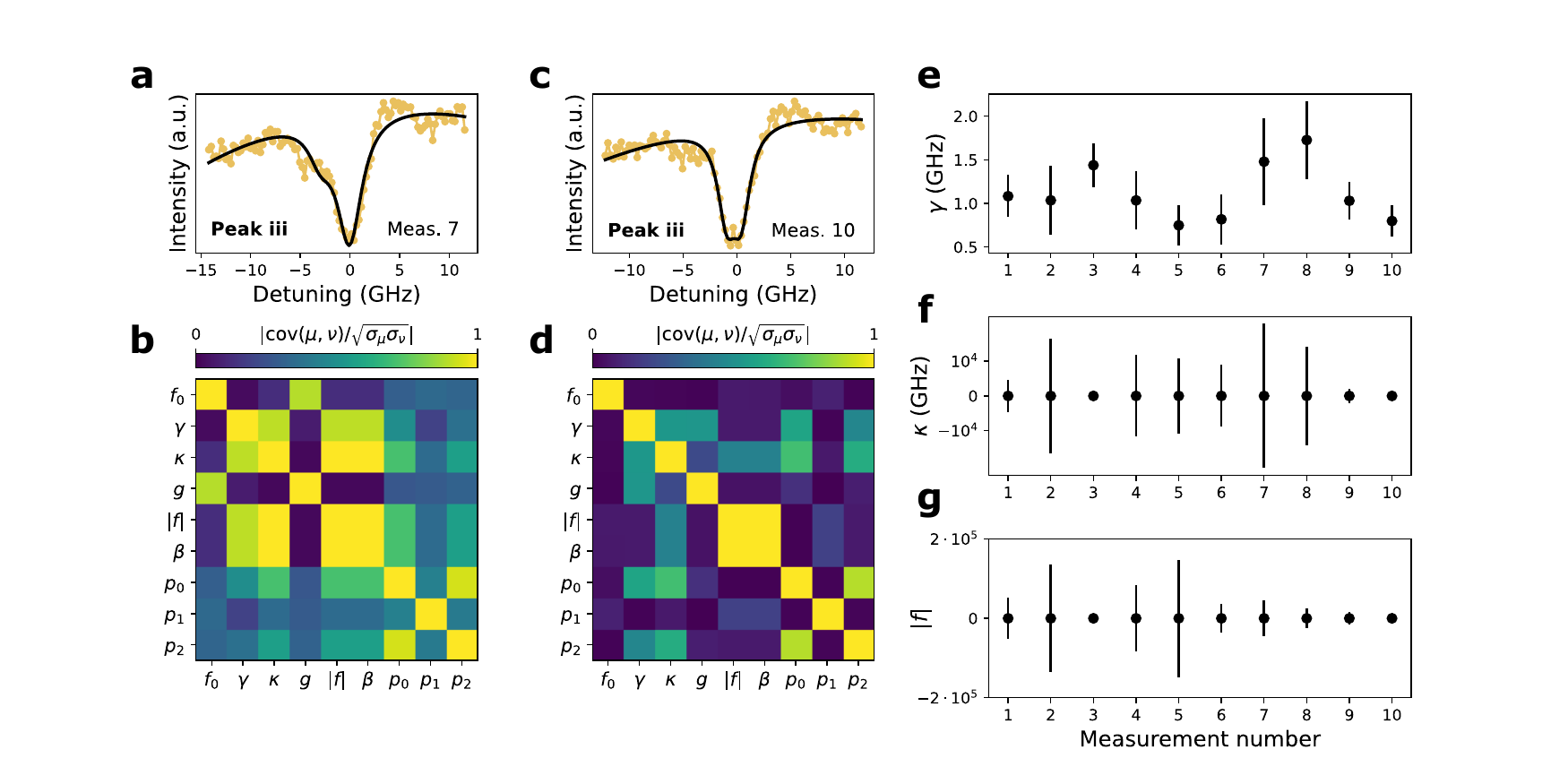}
    \caption{\textbf{Divergence of covariance.}
    \textbf{(a,b)} Raw data with fit for the 7th measurement and associated normalized covariance matrix.
    \textbf{(c,d)} Raw data with fit for the 10th measurement and associated normalized covariance matrix.
    In \textbf{(a)} the fitting included the background feature on the left-hand side as part of the split resulting in a large difference from the constrained fitting.
    The resulting covariance matrix \textbf{(b)} also shows significantly larger covariances between $\kappa$, $|f|$, and $\beta$ as a result of the co-linearity.
    The fit in \textbf{(c)} shows a more reasonable splitting and minimal asymmetry, which results in the relatively strong co-linearity of $|f|$ and $\beta$ (symmetric splitting requires $|f|\cos\beta=0$ which has multiple solutions).
    \textbf{(e,f,g)} Show the fitted values for $\gamma$, $\kappa$, and $|f|$ respectively.
    The co-linearity results in arbitrarily large uncertainty on $\kappa$ and $|f|$, however the estimate of $\gamma$ always retains high confidence.
    The corresponding measurements (7 and 10) show respectively larger/smaller error bars on the co-linear components.
    }
    \label{fig:covariance}
\end{figure*}

Finally, we remark that the convergence of the (un)constrained models is not a result of trivial over-fitting.
The full model (without the weak back-coupling approximation) over-fits the data in the very specific manner of introducing a co-linearity between the forward- and backward-coupling, parameterized by $\kappa$ and $|f|$ respectively.
Such co-linearity, combined with the additional degree of freedom provided by the $\cos\beta$ terms in the numerator of Eq.~\eqref{eq:asymmetric_exact} then allows for nearly any combination of $|f|$ and $\kappa$ to simultaneously optimize the least-square-error objective function.
As a result, the parameters $|f|$, $\beta$, and $\kappa$ typically have large co-variances on each other and estimates of the error on these parameters individually blow up to infinity.
Fortunately, these parameters are independent of the intrinsic loss rate $\gamma$, which is the primary parameter of interest given its relevance to the material loss.
Despite the unbounded errors in the other parameters, estimates of $\gamma$ are determined with high-confidence and remain nearly identical to the quoted values in the main text in nearly all cases.
Examples are shown in Fig.~\ref{fig:covariance}.

\subsection{Quality factor versus coupling distance}
In the main text Figs.~2 and 3, the quality factor as a function of coupling distance is determined to obtain approximate asymptoting values, thus providing an estimate of the intrinsic quality factor $Q_i$.
To perform this fitting we assume that the coupling rate $\kappa$ scales exponentially with the coupling distance $d_c$, such that $\kappa(d_c) = \kappa_0 \exp(-d_c / L)$ where $L$ is some characteristic decay length for the cavity/waveguide evanescent field and $\kappa_0$ is the coupling rate in the limit of $d_c \to 0$.
Thus, we can estimate the total quality factor $Q(d_c)$ by the relation
\begin{align}
    \frac{1}{Q(d_c)} &= \frac{1}{Q_i} + \frac{2}{Q_c},  \nonumber\\
    &\approx \frac{1}{Q_i} + 2 \frac{\kappa_0}{\omega_0}\exp(-d_c / L).
\end{align}
The aggregated quality factor versus coupling distance data was fitted to this equation to estimate the average asymptoting quality factor of $Q_i$.

\clearpage
\section{Simulation of optical resonator modes}
\begin{figure*}[h]
    \centering
    \includegraphics{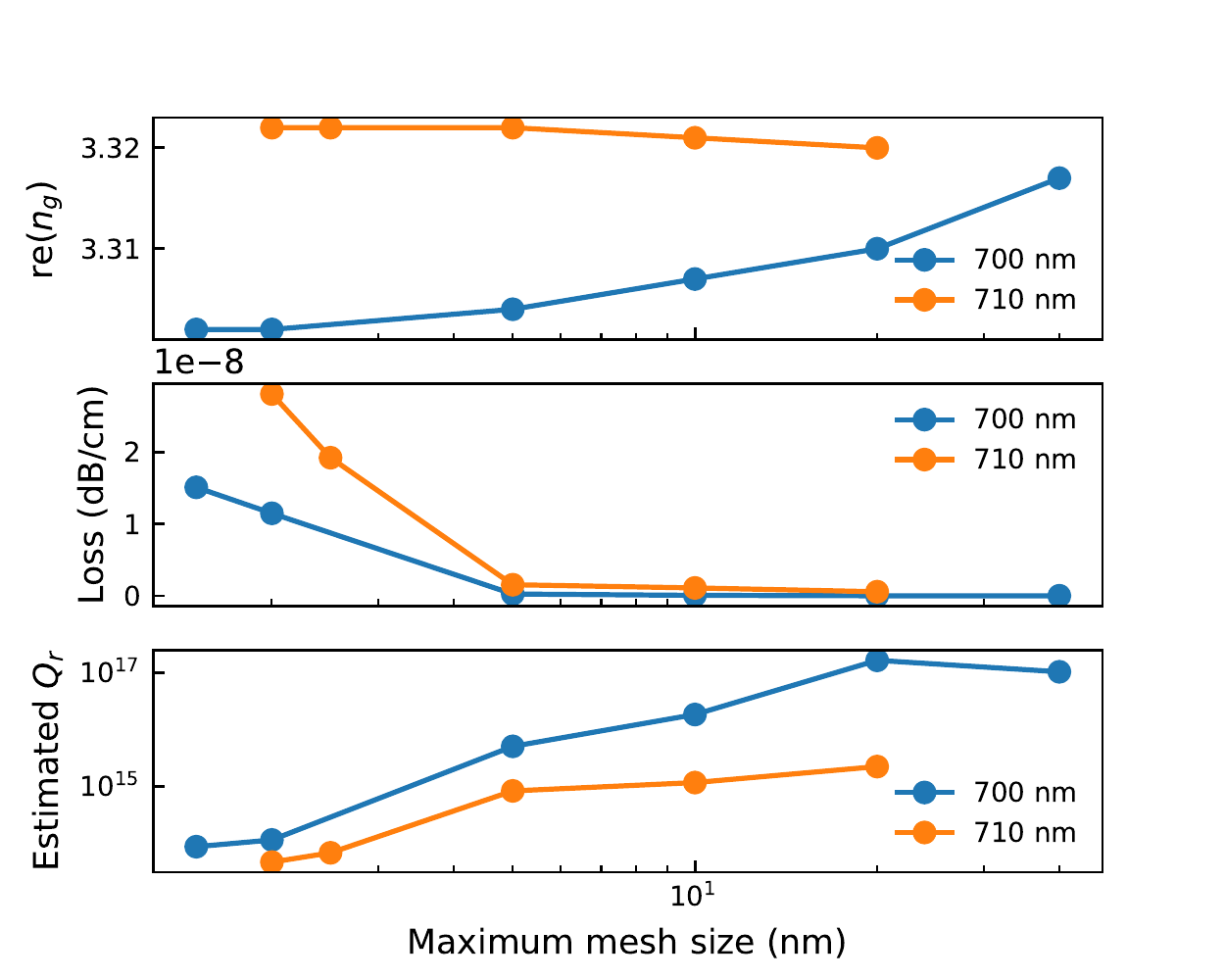}
    \caption{\textbf{Simulation convergence.}
    Convergence of simulation results as a function of minimum mesh size.
    The increase in loss is expected to be a result of the finer mesh size reducing the amount of numerical buffer at the material interfaces.
    Despite this, the expected radiation-limited quality factor $Q_r$ and real part of the group index $n_g$ both converges.
    }
    \label{fig:convergence}
\end{figure*}

The fundamental optical modes of the ring and disk resonators are simulated using a finite-difference eigenmode solver (Ansys Lumerical MODE).
The simulations assume a bending radius appropriate to the specific geometry and assume perfectly matched layer boundary conditions.
A high-resolution mesh is defined at the mode location; convergence of the results is shown in Fig.~\ref{fig:convergence}.
From these simulations we can determine the loss (per length) and group index of the mode.

The simulated loss $\alpha_{[\mathrm{dB/cm}]}$ (given in terms of dB/cm; $\alpha_{[\mathrm{dB/cm}]}\geq 0$) is related to the spatial loss rate $\alpha$ by 
\begin{equation}
    \alpha_{[\mathrm{dB/cm}]} = - 10 \log_{10}\big(e^{-\alpha\cdot (1\,\mathrm{cm})}\big).
\end{equation}
Inverting this one finds
\begin{equation}
    \alpha = - \frac{1}{(1\,\mathrm{cm})} \ln\Big(10^{-\frac{\alpha_{[\mathrm{dB/cm}]}}{10}}\Big).
\end{equation}
The energy in the cavity follows an exponential decay in time
\begin{equation}
    \frac{U(t)}{U_0} = \exp(- \frac{\alpha c t}{n_g})
\end{equation}
where $n_g$ is the group velocity (energy propagates at a speed of $c/n_g$).
The energy decay rate is then given as $\gamma = \alpha c/ n_g$.

\begin{figure*}[h]
    \centering
    \includegraphics{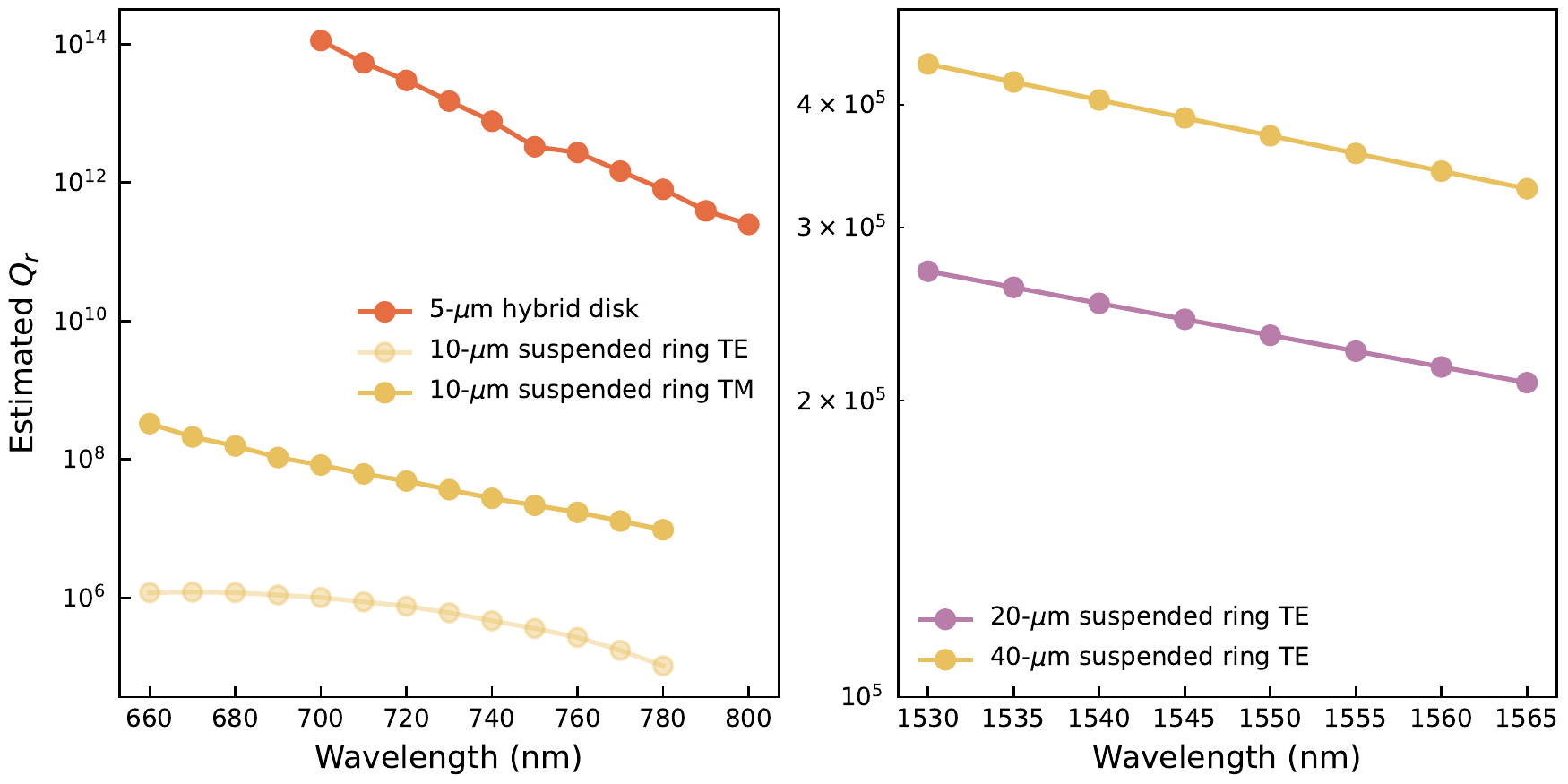}
    \caption{\textbf{Simulated radiation-limited quality factor.}
    Simulated radiation limited quality factors for the designated modes.
    For the optical ring devices, the TM mode is the measured resonance in the main text.
    }
    \label{fig:radiation}
\end{figure*}

The intrinsic quality factor is defined by $Q_i \equiv 2\pi f_0 / \gamma$ where $f_0$ is the center frequency of the cavity resonance.
Expressed in terms of the free-space wavelength $\lambda_0 = c / f_0$ we have
\begin{equation}
    Q_i = \frac{2\pi n_g}{\alpha \lambda}.
\end{equation}
This equation is used to estimate the intrinsic quality factor of the modes as limited by radiation.
The resulting radiation limited quality factors $Q_r$ are shown in Fig.~\ref{fig:radiation}.

\clearpage
\section{High-Q visible wavelength resonances}
\begin{figure*}[h]
    \centering
    \includegraphics[width=5in]{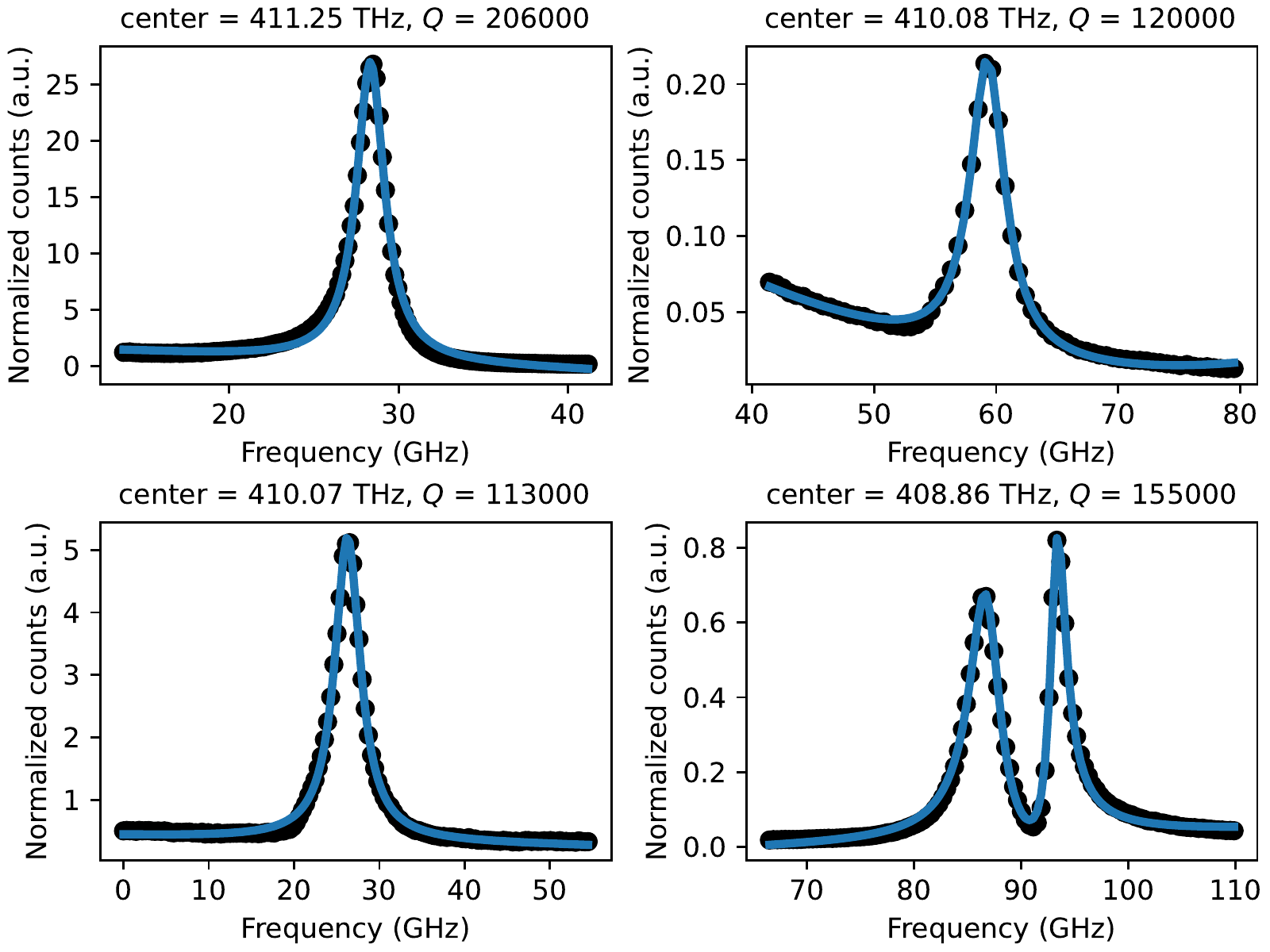}
    \caption{\textbf{High-$Q$ visible-wavelength resonances.} Scanning laser data and fits for visible-wavelength resonances in undoped-BGaP.
    }
    \label{fig:my_label}
\end{figure*}

\clearpage
\section{The BVD model and Electromechanical conversion efficiency calculation}
\begin{figure*}[h]
    \centering
    \includegraphics[width=0.7\textwidth]{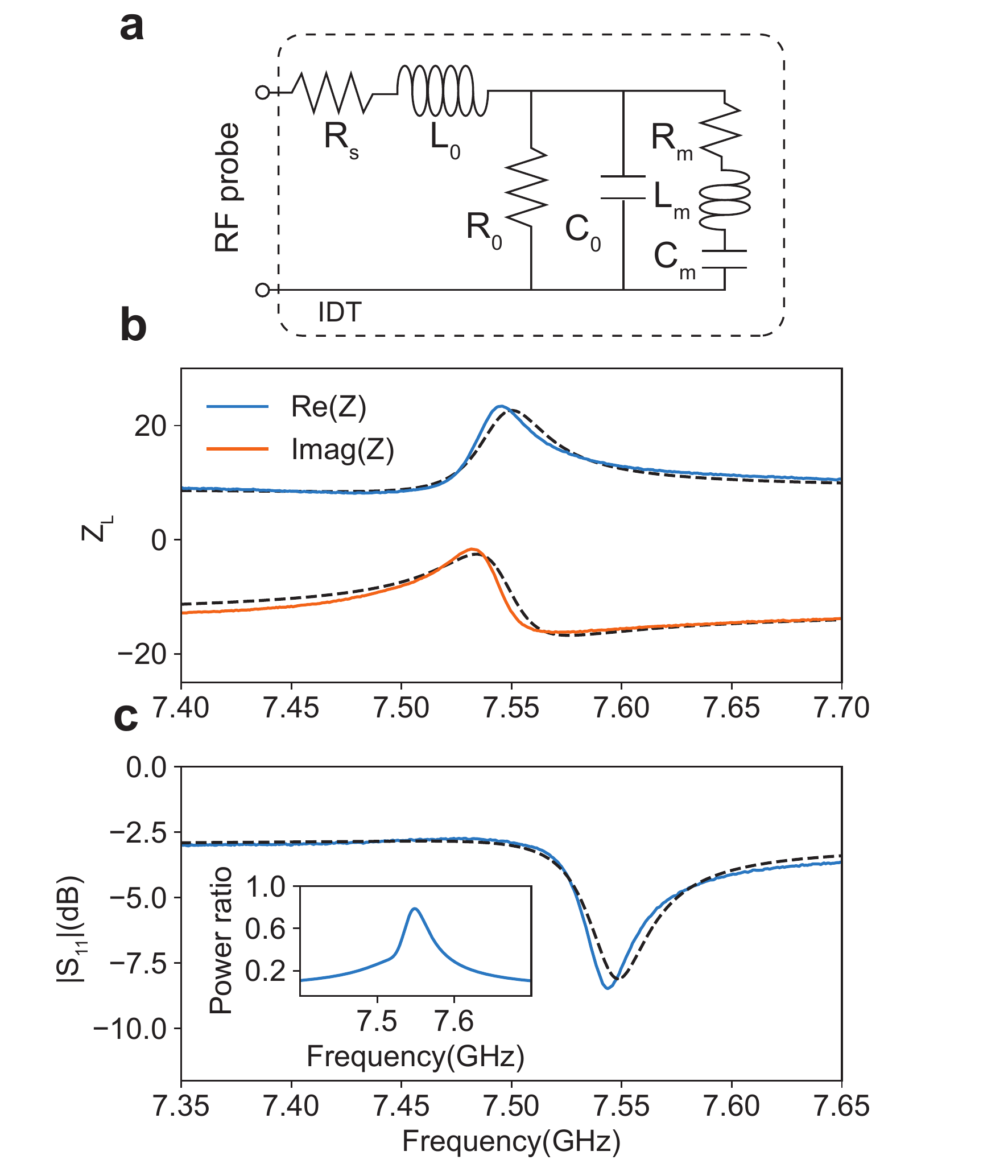}
    \caption{\textbf{BVD model and fitting.}
    The BVD model and the fitting of the effective load impedance. 
    \textbf{(a)} The effective circuit of the BVD model used to fit the $S_{11}$ data. 
    \textbf{(b)} The load impedance calculated from the measured $S_{11}$ data, the dashed lines are the fitted BVD model.
    \textbf{(c)} The $S_{11}$ spectrum and the power ratio on the mechanical lump elements. The dash line is the fitted BVD model. The inset shows the power ratio on the mechanical elements $Y_m$.
    }
    \label{fig:BVD}
\end{figure*}

The RF reflection spectrum is measured on a vector network analyser (VNA)(Keysight N5230C PNA-L), on a cryogenic probe station (Lakeshore CRX-4K). 
The VNA was calibrated at room-temperature and under vacuum, to the tips of the RF probes as the reference using a calibration substrate (GGB Inc. CS-15). 
To extract the IDT's electromechanical conversion efficiency, we use the BVD effective circuit, as shown in Fig.~\ref{fig:BVD}a, to obtain the value of the circuit lump elements and calculate the power load on the motional elements. We can express the load impedance $Z_{\text{L}}$ as:
\begin{equation}
    Z_{\text{L}} =  R_s + j\omega L_0 + \left(\frac{1}{Z_0} + \sum_N Y^{(N)}_m(\omega)\right)^{-1} = \frac{1}{Y_{\text{L}}},
\end{equation}
where
\begin{align}
    &Z_0 = \frac{1}{j\omega C_0 + 1/R_0} \\
    &Y_m(\omega) = \frac{1}{Z_m} = \frac{1}{R_m + j\omega L_N + 1/j\omega C_m}. 
    \end{align}
$R_s$ accounts for the total resistance between the RF probe and the IDT and $L_0$ accounts for the inductive background signal of the free space RF signal. 
The shunt resistor $R_0$ accounts for the power dissipation/loss of the dielectric while the shunt capacitor $C_0$ accounts for the static capacitance of the IDT the electrodes. 
$L_m(\omega)$, $C_m(\omega)$, and $R_m(\omega)$ are the mechanical elements at the target resonance frequency. 
We can combine $C_0$ and $R_0$ as effective load $Z_0$. 
In our case, we only fit one resonance, therefore $N=1$. 
The BVD model fit of $Z_\text{L}$ is shown in Fig.~\ref{fig:BVD}b and the $|S_{11}|$ fit is shown in Fig.~\ref{fig:BVD}c.

After extracting the lump element values, we can calculate the the voltage $V_{Y_m}$ and current $I_{Y_m}$ across $Y_m$ using the Ohm's law.
We find the power ratio of the total input power that is loaded on $Y_m$ to be
\begin{equation}
P_{Y_a} = Re\left[\Tilde{V}_{Y_m}\cdot\Tilde{I}^*_{Y_m}\right].    
\end{equation}
The calculated power ratio on $Y_m$ is shown in the inset of Fig.~\ref{fig:BVD}c,which shows $\sim 80\%$ of the total input power is loaded on $Y_m$ at 4\,K.

\clearpage
\section{Time gating of acoustic transmission measurement}
\begin{figure*}[h]
    \centering
    \includegraphics[width=0.7\textwidth]{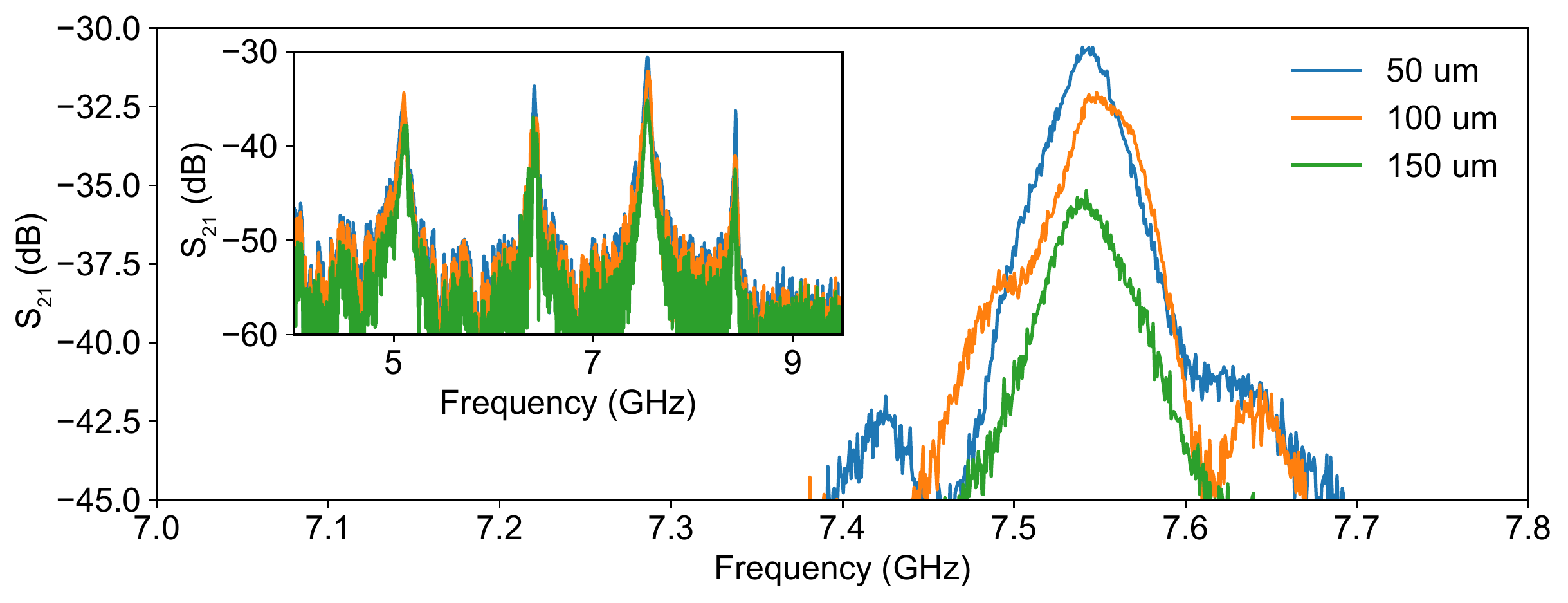}
    \caption{\textbf{S$_{21}$ signal processing.} 
    The frequency domain gated S$_{21}$ signal of the R$_3$ mode. The inset shows the full range of the gated S$_{21}$ spectrum.
    \label{fig:Gate}
    }
\end{figure*}
The measured S$_{21}$ signal includes the acoustic reflections and the free-space RF interference, therefore, we need to perform time-gating to extract the single-pass acoustic transmission signal.
Time-gating of the S$_{21}$ signal is performed by taking an inverse Fourier transform of the measured S$_{21}$, filtering the resulting time-domain signal within a specified window and then Fourier transforming the gated S$_{21}$ signal back to the frequency domain. 
We then take the intensity of the gated signal in the frequency domain as the acoustic transmission of different delay line devices, whose transmission signal was gated between $5 \text{\,ns} < t < 1300 \text{\,ns}$ to remove any signal that is much faster then the acoustic wave. The gated frequency domain signal is shown in Fig. \ref{fig:Gate}.
We did not directly take the intensity of the time-domain impulse signal as the time domain signal contains multiple overlapping acoustic modes.

\clearpage
\section{Heterodyne measurement}
\begin{figure*}[h]
    \centering
    \includegraphics[width=0.7\textwidth]{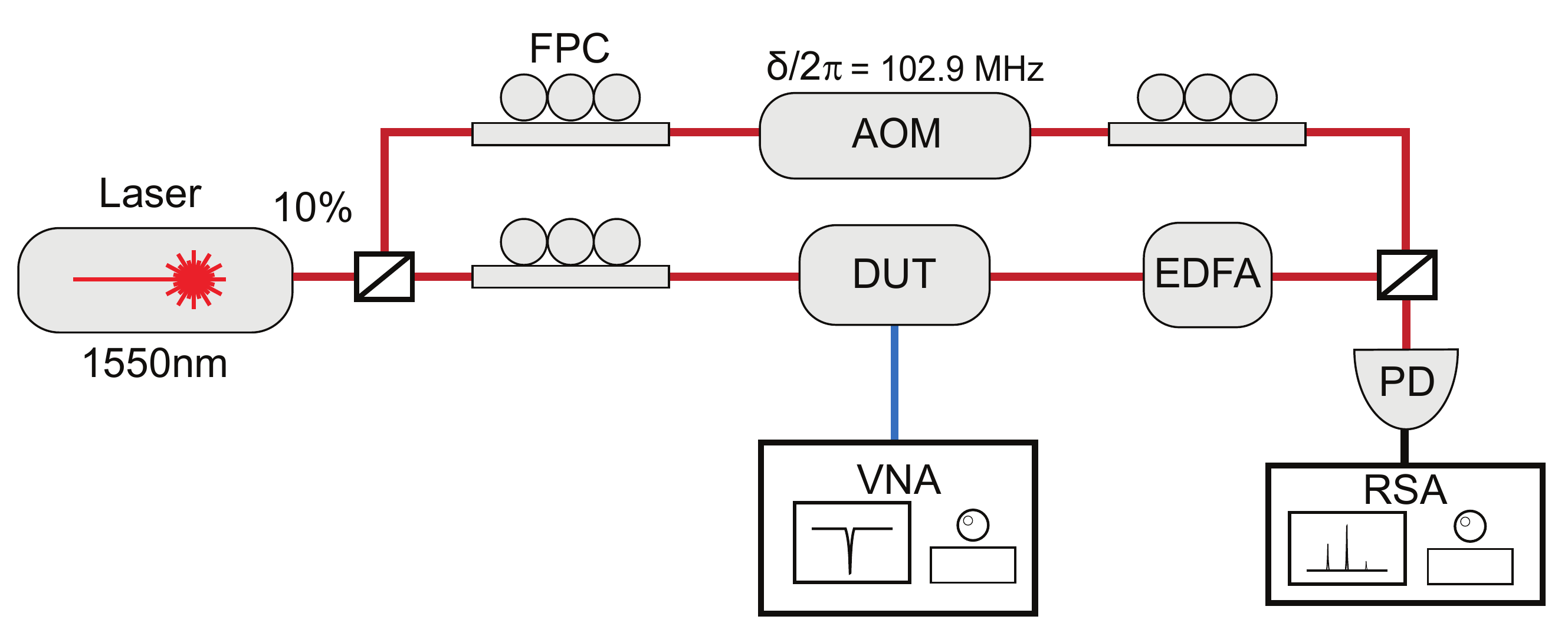}
    \caption{\textbf{Heterodyne measurement setup for the AOFS} FPC: Fiber polarization controller. AOM: Acousto-optical modulator (Brimrose model AMF-100-1550-2FP). EDFA: Erbium-doped fiber amplifier (PriTel LNHP-PMFA-23).  PD: Photodetector (Thorlabs RXM25AF). RSA: Real-time spectrum analyser (Tektronix RSA5100B).
    The input laser is split into two paths. One path goes through the AOM, where the input laser's frequency is up-shifted by $\delta/2\pi = 102.9$ MHz as the reference signal for the heterodyne measurement. The other path goes through the DUT, which is driven by a VNA at $\Omega/2\pi=6.5$ GHz. The two paths then combined and beats together to generate the beating signal with frequency $\Omega \pm \delta$ for Stokes ($+$) and anti-Stokes ($-$) cases.
    \label{fig:heterodyne}
    }
\end{figure*}

\clearpage
\section{Density functional theory}

The density functional calculations were performed using the periodic model approach and plane wave basis set, as implemented in the VASP code \cite{kresse1996prb_dft}.
Plane-wave augmented pseudopotentials were used to represent the core electrons \cite{blochl1994prb_dft}. 
The structure optimization was performed for the 2$\times$2$\times$2 extension of the crystallographic cell (Ge$_{32}$P$_{32}$) using the PBEsol exchange correlation functional \cite{perdew2008prl_dft} in the spin-polarized mode. 
A $\Gamma$-centered 4$\times$4$\times$4 $k$-mesh was used for Brillouin-zone integration. 
The plane wave basis set cutoff was set to 400 eV throughout. 
The total energy was converged to within 10$^{-6}$ eV. 

Prior to density of states (DOS) calculations, the cubic cell parameters were fixed at the value corresponding to 2\% doping level to match the lattice constant of BGaP with the Si lattice constant for consistency with experimental studies. 
Then, the total energy was minimized with respect to the internal atomic coordinates for each defect configuration using the PBEsol and a $\Gamma$-centered 8$\times$8$\times$8 $k$-mesh.

\subsection{Determination of the lattice constant}
Up to four Ga atoms in the Ga$_{32}$P$_{32}$ supercell were replaced with B atoms and the energy of each configuration was minimized with respect to the atomic coordinates and all supercell parameters. 
The pseudo-cubic BGaP lattice constant, derived as the cube root of the supercell volume, decreases with increasing B content (Fig.~\ref{fig:dft_si}) and suggests that BGaP lattice match with that of the bulk silicon is achieved by substituting $\sim$2\% of Ga atoms with B.

\subsection{Interstitial Boron defect structure}
Our calculations suggest that B$_i$ species either bind to the lattice P$^{3-}$ ions or compete with the lattice Ga$^{3+}$ ions for the occupancy of the cation sites, as shown in Fig.~\ref{fig:dft_si}. 
Specifically, a Ga$^{3+}$ ions can be displaced from its lattice site to make room for B so that both of them are coordinated by the lattice P$^{3-}$ ions. 
As discussed in the main text, we found three such B$_i$ configurations that can be described as BP$_{1}$, BP$_{2}$, BP$_{3}$. (Fig.~\ref{fig:dft_si}), i.e., the total number of P atoms coordinating the split B--Ga site remains the same but the number of P--B and P--Ga bonds varies from one to three depending on the configuration.

The energetically favorable boron dimers observed in the calculation can be rationalized in terms of B$^{3+}$ ions having smaller ionic radius than Ga$^{3+}$ ions (0.41 and 0.76 \AA, respectively).
Minimizing the supercell lattice shape and volume in the case of the interstitial B atoms, either in ideal lattice or bound to B$_{\text{Ga}}$, results in the BGaP lattice constant that is similar to or exceeding that of pure GaP as shown in Fig.~\ref{fig:dft_si}.

\subsection{Calculation of the BGaP electronic density of states}
\begin{figure*}[h]
    \centering
    \includegraphics[width=0.5\textwidth]{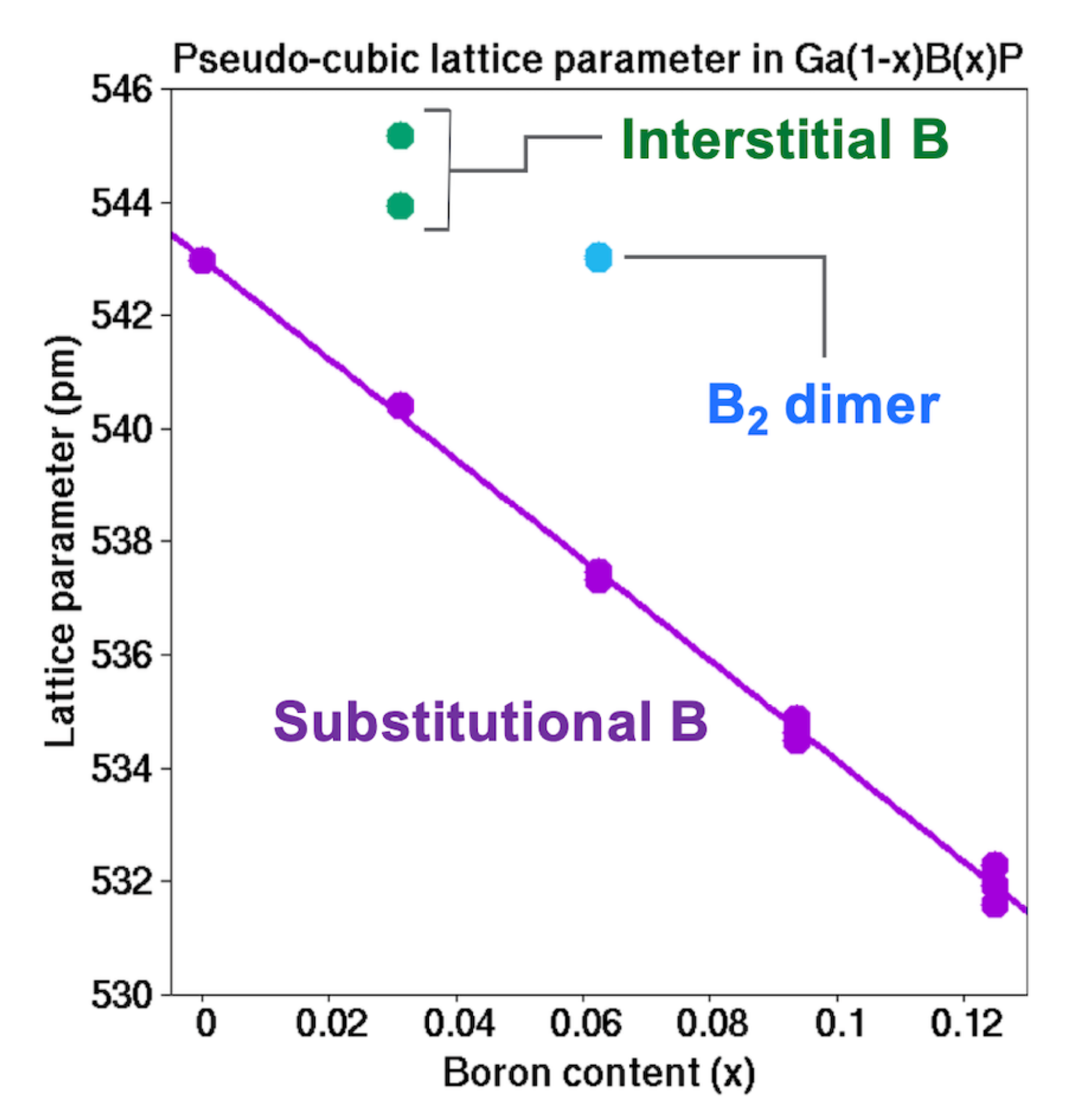}
    \caption{\textbf{Boron content dependence of lattice constant.} 
    Dependence of the pseudo-cubic lattice parameter in GaP:B on the concentration of substitutional B impurities and the corresponding linear fit (purple). 
    Several BGaP configurations were considered for the larger B concentrations. The effect of interstitial B species in the ideal GaP lattice (B$_i$) and those bound to substitutional B (B$_2$) are shown with light green and blue, respectively.
    \label{fig:dft_si}
    }
\end{figure*}

The effect of the interstitial boron defects on the electronic structure are analyzed in terms of the one electron densities of states (DOS). 
It is well known that the generalized gradient approximation exchange correlation functional, such as PBEsol used in this work, underestimates the band gap of semiconductors by as much as a factor of 1.8. 
Accordingly, this deficiency affects the energies of gap states. 
For consistency, we used the cubic supercell and the lattice parameters optimized, as discussed above, for the pure GaP (5.429 \AA) and for GaP:B with 2\% substitutional B defects evaluated using the tend shown in Fig.~\ref{fig:dft_si} (5.412 \AA).

\clearpage

\clearpage
\bibliography{supplement}